\documentclass[a4paper,fleqn,usenatbib]{mnras}

\usepackage{newtxtext,newtxmath}
\usepackage[T1]{fontenc}
\usepackage{ae,aecompl}
\usepackage{graphicx}	
\usepackage{amsmath}	
\usepackage{amssymb}	
\usepackage{multirow}
\newcommand{\avpdf}{$A_V$-PDF}
\newcommand{\avnh}{$A_V{-}n_{\rm H}$}
\newcommand{\caseone}{\mbox{Case-1}}
\newcommand{\casetwo}{\mbox{Case-2}}
\newcommand{\angstrom}{\textup{\AA}}

\title[Atomic and molecular content using distributions]{Simulating the atomic and molecular content of molecular clouds using probability distributions of physical parameters}

\author[T.~G. Bisbas et al.]{
Thomas~G. Bisbas,$^{1,2,3}$\thanks{E-mail: tbisbas@gmail.com (TGB)}
Andreas Schruba,$^{3}$
and Ewine~F. van Dishoeck$^{3,4}$
\\
$^{1}$I.~Physikalisches Institut, Universit\"at zu K\"oln, Z\"ulpicher Stra{\ss}e 77, Germany\\
$^{2}$Department of Physics, Aristotle University of Thessaloniki, GR-54124 Thessaloniki, Greece\\
$^{3}$Max-Planck-Institut f\"ur Extraterrestrische Physik, Giessenbachstrasse 1, D-85748 Garching, Germany\\
$^{4}$Leiden Observatory, Leiden University, P.O. Box 9513, NL-2300 RA Leiden, The Netherlands
}

\date{Accepted XXX. Received YYY; in original form ZZZ}

\pubyear{2018}

\begin{document}
\label{firstpage}
\pagerange{\pageref{firstpage}--\pageref{lastpage}}
\maketitle

\begin{abstract}
Modern observations of the interstellar medium (ISM) in galaxies detect a variety of atomic and molecular species. The goal is to connect these observations to the astrochemical properties of the ISM. 3D hydro-chemical simulations attempt this but due to extreme computational cost, they have to rely on simplified chemical networks and are bound to individual case studies. We present an alternative approach which models the ISM at larger scales by an ensemble of pre-calculated 1D thermo-chemical photodissociation region (PDR) calculations that determine the abundance and excitation of atomic and molecular species. We adopt lognormal distributions of column density (\avpdf s) for which each column density is linked to a volume density as derived by hydrodynamical simulations. We consider two lognormal {\avpdf}s: a diffuse, low density medium with average visual extinction of $\overline{{\rm A}_V}=0.75\,{\rm mag}$ and dispersion of $\sigma=0.5$ and a denser giant molecular cloud with $\overline{{\rm A}_V}=4\,{\rm mag}$ and $\sigma=0.8$. We treat the UV radiation field, cosmic-ray ionization rate and metallicity as free parameters. We find that the low density medium remains fully H{\sc i}- and C{\sc ii}-dominated under all explored conditions. The denser cloud remains almost always molecular (i.e. H$_2$-dominated) while its carbon phase (CO, C{\sc i} and C{\sc ii}) is sensitive to the above free parameters, implying that existing methods of tracing H$_2$-rich gas may require adjustments depending on environment. Our numerical framework can be used to estimate the PDR properties of large ISM regions and quantify trends with different environmental parameters as it is fast, covers wide parameter space, and is flexible for extensions.
\end{abstract}




\section{Introduction}


Astrophysical and cosmological calculations are nowadays becoming increasingly advanced, with the general trend to couple the \mbox{(magneto-)} hydrodynamical evolution of the interstellar medium (ISM) with detailed chemistry in all gas phases. The goal of these new codes is to understand the evolution of molecular clouds and how this links to galaxy evolution \citep[e.g.][]{Inou12,Walc15,Kim17}, to determine how metallicity, ultraviolet radiation and cosmic-rays affect the properties of the ISM \citep[e.g.][]{Giri16,Rich16a,Rich16b}, to identify H$_2$-rich but CO-poor gas in galaxies \citep[e.g.][]{Clar12,Smit14,Bour15}, and to study the star formation process \citep[e.g.][]{Hu17}. Although such codes can simulate the ISM to the point of a direct comparison with observations, they are computationally very demanding and expensive, requiring often an interdisciplinary synergy of groups. 

When it comes to chemistry, a key challenge of such hydro-chemical codes is to determine as detailed as possible the atomic-to-molecular transition and the transition of the different carbon phases \citep[e.g.][]{Glov10,Offn13}. This problem has been the main focus of photodissociation region (PDR) studies over the past 30 years or so \citep{Holl99,Roel07}. In general, PDR codes do not attempt to co-evolve hydrodynamically the ISM. Instead, they apply detailed micro-physics that are sensitive to many user-defined parameters and that together control the overall chemistry, excitation and thermal balance \citep[e.g.][]{vDis88,Ferl98,Bell06,LePe06,Bisb12}. These codes are able to estimate the abundances of hundreds of species and the line emission of various coolants. This information can be used to create synthetic images to be contrasted against real observations \citep[see][for a review]{Hawo17}. However, a fundamental limitation of these codes is that they have to adopt many {\it ad hoc} user-defined parameters describing the ISM, such as the density distribution, gas velocity, the structure of the ultraviolet radiation field and others. These `free' parameters play a crucial role when determining the abundance and emissivities of species and thus affect the physical interpretation of observations.

Coupling hydrodynamics with detailed chemistry has the potential to simulate the ISM quite realistically. Indeed, first promising efforts have been made towards this direction \citep{Bisb15b,Moto15,Hawo15}, although the computational expense is substantial. To reduce the computational cost of hydro-chemical simulations, simpler (and therefore faster) chemical networks have been implemented \citep{Nels97,Glov10,Grass14}. These networks provide an estimate of the local gas temperature, leading to a more accurate estimation of the thermal pressure and therefore the overall ISM hydrodynamics. They are nowadays frequently used to study isolated turbulent molecular clouds in different conditions (from the solar neighbourhood to the Galactic Centre and the distant Universe) and star formation therein when only the most important chemical reactions are accounted for. However, for now extensive parameter studies or detailed simulations of entire galaxies remain prohibitive due to their computational cost.

One important task of chemical calculations is to determine how commonly used methods for tracing H$_2$-rich gas, e.g.\ via CO, depend on various factors. The most important of them are (i) the average ISM metallicity \citep{Pak98}, (ii) the ambient FUV radiation field \citep{Pak98,Wolf03}, (iii) the average temperature, density and dynamical state of the gas \citep{Dick86,Youn91,Brya96,Papa12} and (iv) the CO destruction due to cosmic rays \citep{Bisb15,Bisb17}. These factors are even more important in studies of low-metallicity or high-$z$ galaxies due to their extreme ISM conditions. We thus require predictions of how the partitioning between the carbon phase scales with critical `environmental' parameters.

To connect hydrodynamical simulations, chemical modeling and observations, the visual extinction, $A_V$, probability density function (\avpdf) is considered. At solar metallicity, $A_V$ is related to the total H-nucleus column density, $N_{\rm H}$, through $A_V=N_{\rm H}\cdot6.3\times10^{-22}$  (\citealt{Wein01,Roel07}; see also \citealt{Bohl78,Rach09}). An \avpdf\ is defined as the probability of finding ISM gas within a visual extinction in the range of [$A_V,\,A_V+{\rm d}A_V$]. Various observational studies have determined \avpdf s of Galactic molecular clouds \citep[e.g.][]{Good09,Kain09,Froe10,Abre15,Schn15a,Schn15b,Schn16}, the main characteristic of them being that they consist of a log-normal component for $A_V/\overline{A_V}\lesssim1$ (where $\overline{A_V}$ is the mean observed visual extinction) and a power-law component for $A_V/\overline{A_V}\gtrsim1$. Several theoretical studies explored the nature of these two components \citep{Tass10,Krit11,Burk13,Giri14,Burk15} and concluded that the power-law tail is a result of self-gravity possibly leading to star formation. Other groups have suggested that the power-law tail may be a result of the external pressure to gravitationally unbound entities \citep{Kain11}, that {\avpdf}s of collapsing protoclusters may not exhibit a power-law tail at all \citep{Butl14} or even that an \avpdf\ may be a superposition of various log-normal components \citep{Brun15}. On the other hand, recent studies propose that the distinction between log-normal and power-law components may naturally result from statistical errors connected with the consideration of the last (outer) closed contour \citep{Lomb15,Osse16,Alve17,Kort18} and that all PDFs are in fact power-laws.

Whatever their precise form, it is these \avpdf s that can be used as inputs to estimate the astrochemical properties of the ISM with much simpler models. Indeed, such an approach has been used to connect ensembles of column densities with spectral energy distributions of different cooling lines \citep[e.g.][]{Krum07,Nara08}. One such example is the {\sc despotic} code \citep{Krum14} which uses one-zone models to represent optically thick molecular clouds and calculates line luminosities and cooling rates based on an escape probability formalism. Complementary, \citet{Lero17} use {\sc radex} \citep{vdTa07} to model an ensemble of (one-zone) columns to analyse line emissions in extragalactic observations. These studies move in the direction of constructing a general framework to estimate average ISM properties accounting for complex (unresolved) substructure which is particularly useful when dealing with observations of the turbulent ISM over large spatial scales. However, the previous methods lack of detailed PDR chemistry which could potentially provide better insights into the physical state of the observed ISM.

The motivation of the present work is to further improve this framework to estimate the atomic and molecular content of ISM regions from their substructure by using more detailed and self-consistent thermo-chemical calculations. The proposed method considers user-specified {\avpdf}s as input to parametrise the ISM at large scales and links them with a grid of pre-run 1D thermo-chemical calculations of uniform-density providing average PDR properties, such as abundances of species and line emissivities. To do this, each $A_V$ value is connected to a most probable local H-nucleus number density, $n_{\rm H}$ (and therefore a most probable size of a uniform-density sphere) as determined by hydrodynamical simulations. Each individual PDR is integrated over depth up to this $A_V$ value (also sometimes called $A_V^{\rm tot}$, see \citealt{vDis88}).
The \avpdf\ is the key element to perform chemical modeling of observed ISM regions with complex substructure, without the necessity of performing time-consuming 3D hydro-chemical calculations, and to be able to assess quickly how the ISM characteristics change with critical parameters such as the radiation field ($\chi$), cosmic ray ionization rate ($\zeta_{\rm CR}$) and the metallicity ($Z$). The above is of particular interest in studies of unresolved systems such as extragalactic objects.

In this paper we focus on how the abundances of hydrogen (H$_2$, H{\sc i}) and carbon phase (CO, C{\sc i}, C{\sc ii}) depend on $\chi$, $\zeta_{\rm CR}$ and $Z$ in ISM distributions corresponding to ambient low density atomic gas in galaxies and denser clouds that make up Giant Molecular Clouds. These media are parametrised by lognormal {\avpdf}s (see \S\ref{ssec:approach}). The low density medium has mean extinction along the line-of-sight of the observer of $\overline{A_V} = 0.75$~mag, a dispersion of $\sigma = 0.5$ in logarithmic intervals of $\overline{A_V}$ and mean hydrogen density of $n_{\rm H} \sim 80$~cm$^{-3}$. We will find that such a medium consists predominantly of H{\sc i} and C{\sc ii} (Figure~\ref{fig:cartoon}). The denser molecular cloud has $\overline{A_V} = 4$~mag, $\sigma = 0.8$, $n_{\rm H} \sim 600$~cm$^{-3}$ and consists of H$_2$ and a mixture of CO, C{\sc i}, C{\sc ii}. 

The paper is organized as follows. Section~\ref{sec:ccycle} gives a brief overview on the carbon cycle and the atomic-to-molecular transition. Section~\ref{sec:method} describes the numerical methodology including the PDR calculations. Section~\ref{sec:results} presents the results of our approach. Section~\ref{sec:validation} examines how various other approaches compare to our presented one.  Section~\ref{sec:discussion} discusses the limitations of this methodology and provides directions for further improvements. We conclude in Section~\ref{sec:conclusions}.  Appendix~\ref{sec:app} presents how the atomic-to-molecular transition is affected due to the consideration of the suprathermal formation of CH$^+$, which in turn leads to enhanced CO at low extinctions. Appendix~\ref{sec:app2} presents the here adopted relation between the local volume density and the effective (local) visual extinction. 

In a follow-up paper we will consider the radiative transfer to translate the here derived abundances and excitations to observable emission line fluxes. However, as abundances and excitations are the key elements of modeling the ISM, we will concentrate in this first paper on a detailed investigation of their properties as function of environmental conditions.

\section{The carbon cycle and atomic-to-molecular transition}
\label{sec:ccycle}

\begin{figure}
\centering
\includegraphics[width=0.98\linewidth]{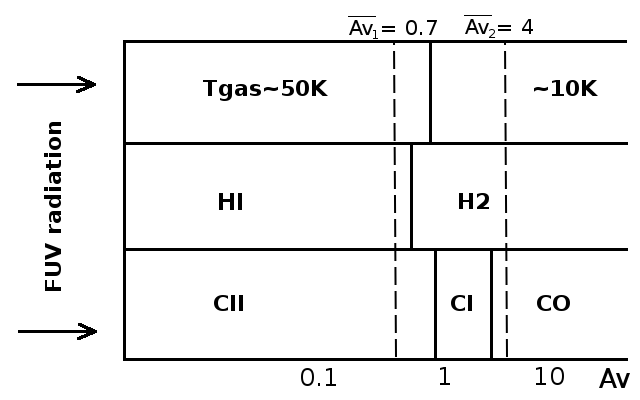}
	\caption{PDR structure for conditions typically found in the Milky Way versus visual (local) extinction, $A_V$. The FUV radiation is assumed to impinge from left-to-right. The extent and actual position of the atomic-to-molecular transition (H{\sc i}-to-H$_2$) and the carbon phase transition (C{\sc ii}-C{\sc i}-CO) depend sensitively on the various parameters examined in this work, such as the density distribution and the depth of each cloud, the FUV radiation field, the cosmic-ray ionization rate, and the ISM metallicity. Varying any of these parameters results in a different stratification of species and different gas temperatures, particularly at low  $A_V$. The above sequence of temperature and abundance structures, however, remains the same. The vertical dashed lines correspond to the mean $A_V$ (denoted as $\overline{A_V}$) for the \caseone\ ($\overline{A_{V1}}$) and \casetwo\ ($\overline{A_{V2}}$) {\avpdf}s described in \S\ref{ssec:approach}. In our 1D PDR models, each cloud is integrated from the edge up to a certain visual extinction along the line-of-sight of the observer (also called $A_V^{\rm tot}$).}
\label{fig:cartoon}
\end{figure}

Figure~\ref{fig:cartoon} illustrates a PDR structure illuminated from one side for conditions typically found in the Milky Way. In order to understand the change in thermal and chemical structure of the ISM, we need to understand when each chemical element transitions from one phase to another. The atomic-to-molecular (H{\sc i}-to-H$_2$) transition is vital for understanding the evolution of the ISM and its chemistry \citep{Jura74,Blac77,vanD86,Glov10,Offn13,Ster14,Bial15b}. This transition is determined and controlled by many different parameters \citep[e.g.][]{Bial17}, the most important of which are the far-ultraviolet radiation \citep[FUV;][]{Jura74,Holl99}, the cosmic-ray ionization rate \citep[e.g.][]{Meij11}, and the gas phase metallicity \citep[e.g.][]{Schr18}. Other parameters such as shock (mechanical) heating \citep{Meij11} and X-ray heating \citep{Malo96,Meij06} may play significant role in special environments, however these are not examined in this work.

Carbon is the most important species for observing the properties of the ISM. It can be found in three major phases: ionized (C{\sc ii}), atomic (C{\sc i}) and molecular in the form of carbon monoxide (CO). All three species are major coolants in the ISM through their spectral line emission \citep{Holl99}. CO is the most abundant molecule after H$_2$ and is frequently used to trace cold H$_2$ gas \citep{Dick86,Solo87,Bola13}, since H$_2$ is a very poor emitter in most ISM conditions. The carbon phase pathway (C{\sc ii}/C{\sc i}/CO) is connected with different ISM evolutionary stages \citep[e.g.][]{vDis88,Ster95,Beut14,Walc15}, and studying it is thus of significant importance. 

In the initial stages of the formation of clouds, the ISM is found in a diffuse and ionized state, thus rich in C{\sc ii} and bright in the [C{\sc ii}] $158\,\mu$m fine-structure line \citep{Roel06,Nord16}. Here, hydrogen is mainly found in atomic form (H{\sc i}), unless a source of Lyman-$\alpha$ photons is present ($h\nu>13.6\,{\rm eV}$, i.e. an OB star or cluster of massive stars)
which ionizes the ISM and creates H{\sc ii} regions. Turbulence and self-gravity can increase locally the density of the ISM leading to larger total column densities \citep{Elme04,Henn12}. This creates conditions in which C{\sc ii} recombines forming C{\sc i} and subsequently CO. Due to its higher abundance and self-shielding, hydrogen is transformed earlier than carbon (i.e. at lower visual extinctions) and forms H$_2$ molecules. CO can also self-shield, but to a lesser degree than H$_2$, so dust absorption is also important. Thus, if the ISM density increases even more, to conditions typically found in giant molecular clouds, the dust shields the propagation of the FUV ($6<h\nu<13.6\,{\rm eV}$) radiation and this creates vast regions of molecular gas rich in CO and other molecules \citep[e.g.][]{vDis88,Berg04,Lang10,Glov11}. 

Cosmic-rays penetrate the ISM even at high column densities and change their chemistry in places where (external) FUV radiation cannot reach \citep[see][for a review]{Stro07,Gren15}. The effect of elevated cosmic-ray energy densities on the chemistry related to carbon phases and the atomic-to-molecular transition was studied numerically by \citet{Meij11} and analytically by \citet{Bial15}. \citet{Bisb15, Bisb17} investigated in detail the consequences on the traceability of molecular gas in star-forming galaxies at low and high redshift, since higher $\zeta_{\rm CR}$ values scale proportionally to the star formation rate \citep{Papa10}. As the $\zeta_{\rm CR}$ ionization rate increases, cosmic-rays produce a high amount of He$^+$ which then reacts with CO creating C{\sc ii}. The latter then quickly recombines with free electrons creating C{\sc i}. At the same time, the H$_2$ molecule remains remarkably unaffected for densities expected in diffuse molecular clouds. This effect of cosmic-ray induced CO destruction has the potential to create vast amounts of CO-poor H$_2$ gas \citep{Bisb15}, much higher than those created due to the presence of FUV radiation. The term `CO-poor' is used to describe the molecular gas state in which the CO abundance with respect to H$_2$ is significantly lower than normally expected, whether caused by enhanced cosmic rays or any other parameter \citep{vDis92}.

Low metallicity gas as typically found in dwarf galaxies \citep[e.g.][]{Star97,Tafe10,Requ16,Pine17}, the outer parts of large galaxies \citep[$R\gtrsim15\,{\rm kpc}$,][]{Bals11,Hayd14} and galaxies at high redshift \citep[e.g.][]{Carn18} have low dust-to-gas ratios which reduce the H$_2$ formation rate (which happens on the surfaces of dust grains) as well as the shielding of CO photodissociation by dust.  In addition, the reduced C and O abundances imply that the CO column builds up more slowly so that its self-shielding sets in later. This has profound effects on all chemical processes of these systems  \citep{Malo88,Wolf95,Pak98,Madd06,Bial15} and modifies the stratification of PDR species (Fig.~\ref{fig:cartoon}). Observations of low metallicity clouds show that CO is restricted to small, well shielded dense gas peaks surrounded by H{\sc i}- and H$_2$-rich gas \citep[e.g.][]{Schr17}, and that CO directly scales with the line-of-sight extinction $A_V$  \citep[][]{Lee15,Lee18}.

\section{Numerical method}
\label{sec:method}

\subsection{PDR astrochemical calculations}
\label{ssec:supr}
For the purposes of this work, the publicly available code {\sc 3d-pdr}\footnote{https://uclchem.github.io/3dpdr.html} is used, which treats the astrochemistry of photodissociation regions \citep{Bisb12}. Although the code is able to perform PDR calculations for 3D density distributions, here we apply it to ensembles of 1D columns. The gas column is illuminated from one side by a plane-parallel  ultraviolet (UV) radiation field $\chi$ \citep[normalized to the spectral shape of][]{Drai78} with UV radiation decreasing with depth into the cloud. We consider UV photons belonging to the Lyman-Werner band ($11.2-13.6\,{\rm eV}$) which can dissociate molecules but do not ionize hydrogen. The attenuated UV radiation is a function of the optical depth $\tau$ and is given by $\chi=\chi_0\exp(-\omega_{\lambda}\tau)$ with $\omega_{\lambda}=3.02$ at $\lambda=1000\,\angstrom$ \citep[see][]{Roel07}. The code performs full thermal calculations by balancing various heating and cooling processes as function of depth along the column, and adopting an escape probability approach determines the local gas temperature as a self-consistent solution \citep[see][for full details]{Bisb12,Bisb17}. The abundances of species and their line emissivities are then calculated for the local (attenuated) UV field and gas density. Here, a subset of the UMIST2012 network is used \citep{McEl13} consisting of 33 species and 330 reactions. 

In this work, CO freeze-out on dust grains is not included. However, the route of suprathermal formation of CO via CH$^+$ is included, following the methodology described by \citet{Viss09}.  This route is found to give a better agreement between the column densities of H$_2$ and CO at low visual extinctions. \citet{Fede96} argued that the observed enhancement of CO column densities at low visual extinctions ($A_V<1\,{\rm mag}$) is possibly due to non-thermal motions between ions and neutrals as a result of Alfv\'en waves interacting with the cloud in its outermost parts, thus helping to overcome the energy barrier of the reaction
\begin{eqnarray}
{\rm C}^++{\rm H}_2\longrightarrow {\rm CH}^++{\rm H}.
\label{reac:sup}
\end{eqnarray}
This interaction increases the gas (kinetic) temperature by the amount of $\mu v_{A}^2/3k_{\rm B}$, where $\mu$ is the reduced mass of reactants from all ion-neutral reactions, $v_A$ is the Alfv\'en speed and $k_{\rm B}$ is the Boltzmann constant. This increment in gas temperature is able to change the pathway normally followed to form CO, the latter being formed now due to reactions between CH$^+$ and O in addition to C$^+$ and OH \citep{Fede96}. Interestingly, such a turnover in the formation of CO via the OH channel has also been found to result from elevated cosmic-ray energy densities \citep{Bisb17,Bial15}. This means that cosmic-rays can initiate an alternative chemical pathway of CO formation. In addition, the above reaction shifts the H{\sc i}-to-H$_2$ transition at higher $A_V$, since it dissociates H$_2$ in the outermost parts of the cloud where it could otherwise remain molecular, i.e. if the energy barrier cannot be overcome (see Appendix~\ref{sec:app}).

\begin{figure}
\centering
\includegraphics[width=0.98\linewidth]{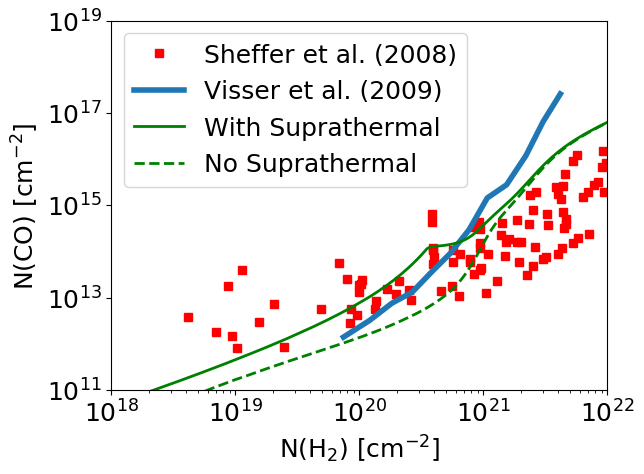}
\caption{Correlation of $^{12}$CO and H$_2$ column densities. Red squares show the observations by \citet{Shef08}. The blue solid line is the mean value from 100 different isothermal PDR calculations performed by \citet{Viss09}. Green lines show the results from a simple full thermal balance calculation of a gas column with $n_{\rm H}=100\,{\rm cm}^{-3}$ irradiated by $\chi/\chi_0=1$ using {\sc 3d-pdr} with (solid) and without (dashed) suprathermal formation (see text for full details). The small kink at N(H$_2$)$\sim\!4\times10^{20}\,{\rm cm}^{-2}$ is when suprathermal formation of CO via CH$^+$ ceases to be important (corresponding to $A_V\sim0.7\,{\rm mag}$).}
\label{fig:sup}
\end{figure}

To demonstrate how the suprathermal formation of CO via CH$^+$ operates, we have performed a full thermochemical PDR calculation of an one-dimensional uniform density distribution with $n_{\rm H}=100\,{\rm cm}^{-3}$ interacting with a plane-parallel radiation field with strength $\chi/\chi_0=1$. Here, the cosmic-ray ionization rate is taken to be $\zeta_{\rm CR}=1.3\times10^{-17}\,{\rm s}^{-1}$, the Alfv\'en speed is taken to be $v_{A}=3.3\,{\rm km}\,{\rm s}^{-1}$  and the calculation is terminated once steady-state thermal balance has been reached. The initial gas phase elemental abundances considered (normalized to hydrogen), where He~$=0.1$, C$^+=1.4\times10^{-4}$ and O~$=3.1\times10^{-4}$. All aforementioned values are chosen to allow for a direct comparison with the results of \citet{Viss09}. 

Figure~\ref{fig:sup} shows the resultant correlation of $N$(H$_2$) and $N$(CO) when the suprathermal formation of CO via CH$^+$ has been considered (green solid line) and when it has not (dashed line). As can be seen, $N$(CO) is approximately one order of magnitude higher in the first case when $N$(H$_2$)$\lesssim10^{21}\,{\rm cm}^{-2}$, the latter of which corresponds to a visual extinction of $A_V\sim1.5\,{\rm mag}$. The red squares are observations taken by \citet{Shef08} in Galactic diffuse molecular clouds. The blue solid line is the correlation estimated by \citet{Viss09} obtained by averaging 100 isothermal PDR calculations. Since a significant fraction of the ISM can be in this low column density phase, it is important to treat this region accurately.

\subsection{Abundances of species for given {\avpdf}s}
\label{ssec:approach}

The aim of this work is to construct a method that determines the average chemical properties of the ISM, i.e. the abundances and excitation of species, by using as input probability density functions (PDFs) of the main physical parameters that describe interstellar clouds. The most fundamental parameter for this work is the total hydrogen column density $N_{\rm H}$-PDF. Hereafter, this parameter will be referred to with the more commonly used relation of {\avpdf}, where $A_V$ is the extinction related to $N_{\rm H}$ through $A_V=N_{\rm H}\cdot6.3\times10^{-22}$ for solar metallicity galactic clouds \citep{Wein01,Roel07}. When modelling lower metallicity regimes, the same $N_{\rm H}$-PDF is adopted but the associated $A_V$ varies proportional to the metallicity, $Z$. In our 1D-PDR models, the visual extinction in the \avpdf\ function is taken  to be the extinction integrated along the line-of-sight of the observer. The latter is also denoted as $A_V^{\rm tot}$.

\begin{figure}
\centering
\includegraphics[width=0.98\linewidth]{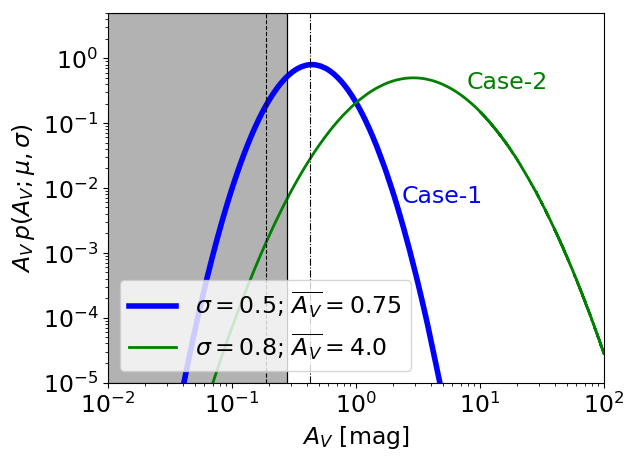}
\caption{Hypothetical {\avpdf}s for ISM corresponding to a diffuse, atomic-dominated medium (blue solid line with $\overline{A_V}=0.75$~mag and $\sigma=0.5$) and a denser giant molecular cloud (green solid line with $\overline{A_V}=4$~mag and $\sigma=0.8$). The shadowed region shows the values of $A_V$ which are not connected with modelled $n_{\rm H}$ densities in this work as estimated from the `average' \avnh\ (see Fig.~\ref{fig:avnh} and Appendix~\ref{sec:app2}). The vertical dashed and dot-dashed lines correspond to the `lower' and `upper' bounds of $A_V$, respectively. These formulae do not include any power-law tail that is potentially expected to appear in a star-forming region described by the \casetwo\ PDF (see text).}
\label{fig:testpdf}
\end{figure}

As a first approach towards this aim, simple lognormal \avpdf\ functions of the following form are considered:
\begin{eqnarray}
\label{eqn:pdf}
{\rm PDF}(A_V;\mu,\sigma)=\frac{1}{A_V \sigma\sqrt{2\pi}}\exp\left[-\frac{(\ln A_V-\mu)^2}{2\sigma^2}\right],
\end{eqnarray}
where $\mu\equiv\overline{\ln A_V}$ and $\sigma$ are the mean and standard deviation of the variable's natural logarithm, respectively. Three different statistical characteristic values can be then defined. The \emph{mean} value, $M$, which corresponds to the average value of all $A_V$ (i.e. $\overline{A_V}$), defined as 
\begin{eqnarray}
\label{eq:var1}
M\equiv\exp\left(\mu+\frac{\sigma^2}{2}\right)=\overline{A_V},
\end{eqnarray}
the \emph{median} value, $m$, which corresponds to the value of $A_V$ that divides the PDF area in two equal parts, defined as
\begin{eqnarray}
\label{eq:var2}
m\equiv\exp(\mu)=\overline{A_V}\exp\left(-\frac{\sigma^2}{2}\right),
\end{eqnarray}
and the \emph{mode}, ${\cal M}$, which corresponds to the peak value of the PDF (where $dp/dA_V=0$), defined as
\begin{eqnarray}
\label{eq:var3}
{\cal M}\equiv\exp(\mu-\sigma^2)=\overline{A_V}\exp\left(-\frac{3\sigma^2}{2}\right).
\end{eqnarray}
By adopting appropriate values of $\overline{A_V}$ and $\sigma$, two different {\avpdf}s are created corresponding to a hypothetical low density, atomic-dominated medium (with $\overline{A_V}=M=0.75\:{\rm mag}$ and $\sigma=0.5$ giving $m=0.66$ and ${\cal M}=0.52$; hereafter `\caseone') and a denser giant molecular cloud (with $\overline{A_V}=M=4\:{\rm mag}$ and $\sigma=0.8$ giving $m=2.90$ and ${\cal M}=1.53$; hereafter `\casetwo'). We note that we do not attempt to model any specific region reported in observations and that we ignore any power-law tail that may appear in a region similar to the {\casetwo} (although its contribution can be considered by the presented method, as discussed in \S\ref{ssec:case3}). Instead, the mean and the width are chosen to roughly mimic typical structures of the ISM to test the proposed method and determine trends with physical parameters. Such PDFs have been found in the local ISM \citep[e.g.][]{Kain09}. The shapes of these PDFs are illustrated in Fig.~\ref{fig:testpdf}. 

To estimate the average column-integrated fractional abundances, $f_{\rm sp}$, of each different species (sp), the following formula is applied
\begin{eqnarray}
\label{eqn:fsp}
f_{\rm sp}=\frac{\sum_{i=1}^qN_i({\rm sp})\cdot {\rm PDF}_i}{\sum_{i=1}^qN_i({\rm H,tot})\cdot {\rm PDF}_i},
\end{eqnarray}
where $N_i$ is the column density multiplied by the frequency ${\rm PDF}_i$, which is determined by the \avpdf\ function given by Eqn.~\ref{eqn:pdf} and $N({\rm H,tot})=N({\rm H\textsc{i}})+2N({\rm H_2})$ is the total hydrogen column density. The summation over the index $i$ considers the PDR calculations at a linear grid of $A_V$ values.

\begin{figure}
\centering
\includegraphics[width=0.98\linewidth]{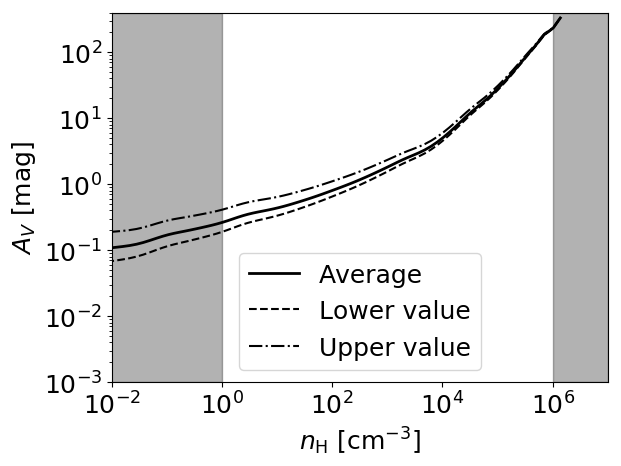}
\includegraphics[width=0.98\linewidth]{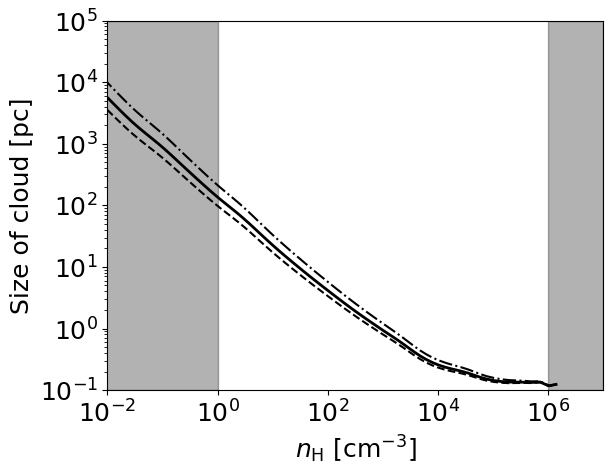}
	\caption{{\it Top panel}: The `default' \avnh\ relation described in Appendix~\ref{sec:app2} (solid black line) along with its adopted upper (dot-dashed line) and lower (dashed line) bounds. This relation connects the most probable value of $A_V$ with $n_{\rm H}$ based on hydrodynamical simulations from pc- to kpc-scales. The upper and lower bounds represent the spread in $A_V$ for constant $n_{\rm H}$ as dictated by hydrodynamical simulations. These bounds decrease by increasing $n_{\rm H}$. In this work, each \avnh\ curve is treated independently. {\it Bottom panel}: the hypothetical size of a cloud (extent along the line-of-sight) for a given H-nucleus number density for the above \avnh\ relations. In both panels, shaded regions correspond to extreme densities whose chemistry is not modeled here. For any $A_V$ outside the bounds, we adopt the chemical composition of the respective boundary PDR calculation.}
\label{fig:avnh}
\end{figure}

To estimate the column density, $N$, a relation connecting the H-nucleus number density, $n_{\rm H}$, with the most probable value of visual extinction, $A_V$, is required. It is reasonable to expect that low $n_{\rm H}$ are most likely found at low $A_V$, therefore being more strongly affected by UV radiation. On the contrary, high $n_{\rm H}$ are expected at high $A_V$. Three-dimensional hydrodynamical simulations from parsec \citep{Glov10,Offn13} to kiloparsec scales \citep{VanL13,Seif17,Safr17,Gong18} remarkably demonstrate this correlation. In this work, the `default' \avnh\ relation is an average over four different such relations found in the literature (see Appendix~\ref{sec:app2} for details). In the upper panel of Fig.~\ref{fig:avnh}, the default \avnh\ relation is plotted as solid line along with the adopted upper (dot-dashed line) and lower bounds (dashed line) to mimic the relations derived from hydrodynamical simulations. In the following, each upper and lower bound is treated independently as an additional {\avnh} relation. This is to demonstrate how the estimated abundances of species depend on the choice of such relation.

By relating the line of sight integrated visual extinction, $A_V$, with the number density $n_{\rm H}$, we are in fact connecting each cloud with a specific uniform-density PDR calculation. Strictly speaking, this is an approximation since $A_V$ is a local value in the simulations used to derive this relation. Our constant density clouds are used as a simplified proof of concept for the statistics presented here. A more realistic assumption would be to use a given \avnh\ relation as an input \emph{directly} to the PDR code, so that the edge of the cloud has a lower density than the center,  and calculate the thermal balance and abundances of species accordingly. The latter method is to be developed in a subsequent paper. 

The \avnh\ relations described above imply an average size, $L$, corresponding to the `depth' or the `line-of-sight extent' of a hypothetical uniform density sphere of $L=A_V/6.3\times10^{-22}n_{\rm H}$. The bottom panel of Fig.~\ref{fig:avnh} shows the scaling between $L$ and $n_{\rm H}$. The mode of the `\caseone' \avpdf\ corresponds to a medium with $n_{\rm H}\sim20\,{\rm cm}^{-3}$, $L\sim13\,{\rm pc}$ and $M\sim4.5\times10^3\,{\rm M}_{\odot}$ and the mode of the `\casetwo' corresponds to $n_{\rm H}\sim600\,{\rm cm}^{-3}$, $L\sim1.3\,{\rm pc}$ and $M\sim136\,{\rm M}_{\odot}$. 

In our analysis we consider two density limits: a minimum of $n_{\rm H}=1\,{\rm cm}^{-3}$ and a maximum of $n_{\rm H}=10^6\,{\rm cm}^{-3}$, below and above which no PDR modelling is performed. Regions corresponding to densities (and thus $A_V$ values) outside these bounds are shown as grey shaded regions in Figs.~\ref{fig:testpdf} and Fig.~\ref{fig:avnh}. For these regions we adopt the 1D-PDR calculation performed at the respective density limit and integrate the PDR model to the total $A_V$ of that cloud. For $A_V>30\,{\rm mag}$ (and consequently for the models with $n_{\rm H}\gtrsim10^5\,{\rm cm}^{-3}$), we adopt the chemical composition calculated at the maximum extinction of $A_V=30\,{\rm mag}$ but integrate the PDR models to the $A_V$ as dictated by the \avnh\ relation. This inserts negligible errors both due to the fact that the UV radiation has been severely attenuated and thus chemistry remains unchanged, and also because the probability of such a high $A_V$ is typically negligible ($\lesssim10^{-4}$ for the \casetwo\ {\avpdf}).

A large grid of constant density simulations is then constructed (see \S\ref{sec:results}) and a look-up table is used to recall a particular PDR simulation and to calculate the column densities of CO, C{\sc i}, C{\sc ii}, H{\sc i} and H$_2$  (the species examined in this paper), as well as the total H-nucleus column density, up to that value of visual extinction. Equation~\ref{eqn:fsp} is then applied for both \avpdf\ cases. Chemical iterations are converged once equilibrium is reached. {\sc 3d-pdr} performs these calculations for a chemical time of $8\,{\rm Myr}$, although equilibrium is typically reached after the first couple Myrs. 

\subsection{Abundances of individual uniform-density PDRs}
\label{ssec:ref}

\begin{table}
	\centering
	\caption{Initial gas-phase abundance of species used for the grid of PDR simulations. The gas-to-dust ratio is assumed to be 100.}
	\label{tab:abun}
	\begin{tabular}{lclc}
		\hline
		\hline
		H     & $4.00\times10^{-1}$ & C$^+$ & $1.00\times10^{-4}$\\
		H$_2$ & $3.00\times10^{-1}$ & O & $3.00\times10^{-4}$\\
		He    & $1.00\times10^{-1}$ &  & \\
		\hline
	\end{tabular}
\end{table}

As reference for our chemical modeling, we consider the following set of one-dimensional, uniform density PDR calculations. We consider an H-nucleus number density that spans 6 orders of magnitude\footnote{Densities higher than $10^6\,{\rm cm}^{-3}$ correspond to very small probabilities in both PDF cases explored and, as we consider only relatively low critical density molecules, are not considered further.} ($n_{\rm H}=10^0-10^6\,{\rm cm}^{-3}$) at a resolution of $0.01$ per logarithmic dex (therefore $q=600$) and assume a maximum visual extinction of $A_V^{\rm max}=30\,{\rm mag}$ at all times.
The other `environmental' parameters are kept fixed at a plane-parallel FUV radiation field with strength $\chi/\chi_0=1$ \citep{Drai78}, a cosmic-ray ionization rate per H~atom of $\zeta_{\rm CR}=10^{-16}\,{\rm s}^{-1}$ representing the average value observed in the Milky Way \citep{Dalg06, Indr15, Neuf17}, and adopting solar metallicity. The column density of each species is derived by integrating the abundance at each depth into the cloud up to the $A_V^{\rm tot}$ as sampled by the \avpdf\ of either \caseone\ or \casetwo.  The initial gas-phase chemical abundances are listed in Table~\ref{tab:abun}. These abundances take into account that some fraction of carbon and oxygen is locked up in solids.

\begin{figure}
\centering
\includegraphics[width=0.95\linewidth]{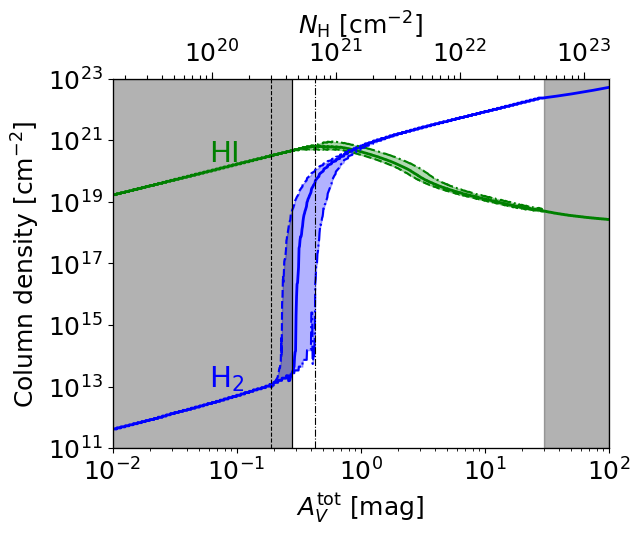}
\includegraphics[width=0.95\linewidth]{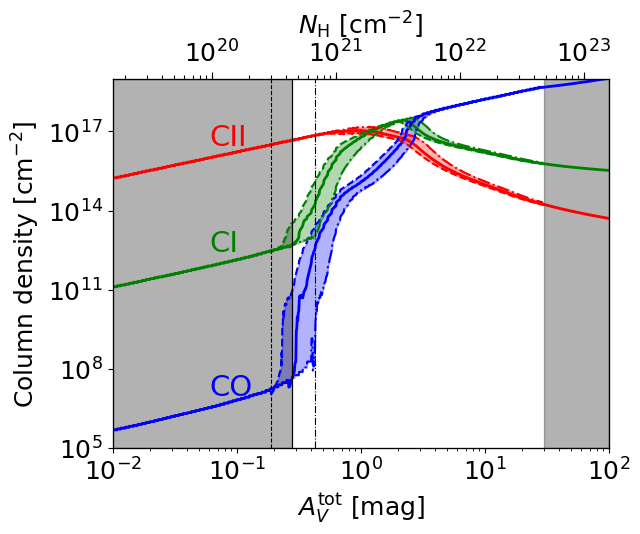}
	\caption{Column densities of H{\sc i} and H$_2$ (top panel) and C{\sc ii}, C{\sc i}, CO (bottom panel) of individual one-dimensional PDRs with volume density $n_{\rm H}$ and total extinction $A_V^{\rm tot}$, following the `average' \avnh\ relation (solid lines). The shaded regions on the left mark the assumed extent of the \avnh\ relation for a constant $n_{\rm H}$, as shown in Fig.~\ref{fig:avnh}. In particular, the dot-dashed lines refer to the `upper bound' and dashed lines to the `lower bound'. Each PDR at constant $n_{\rm H}$ is integrated up to the depth $L$ of each cloud (bottom panel of Fig.~\ref{fig:avnh}). The shaded regions on the right mark the area where we adopt the chemical composition calculated at $A_V=30\,{\rm mag}$ but integrate up to the $A_V$ as dictated by the \avnh\ relation. The `reference ISM' simulations (see \S\ref{ssec:ref}) assumes a FUV radiation field of $\chi/\chi_0=1$, a cosmic-ray ionization rate of $\zeta_{\rm CR}=10^{-16}\,{\rm s}^{-1}$, and solar metallicity.
	}
\label{fig:cdcycle}
\end{figure}

Figure~\ref{fig:cdcycle} shows the column densities of H{\sc i} and H$_2$ (upper panel) and carbon phase (lower panel) for individual uniform density, one-dimensional PDRs as a function of $A_V^{\rm tot}$ when adopting the `average' \avnh\ relation (solid lines) or the upper or lower bounds (dot-dashed or dashed lines). For these PDRs, the atomic-to-molecular transition occurs at $A_V\sim0.9\,{\rm mag}$ (we note that we define this transition by $N_\textsc{Hi} = N_{\rm H_2}$) which corresponds to a column with density $n_{\rm H}\sim130\,{\rm cm}^{-3}$ and length $L\sim3.2\,{\rm pc}$. The PDRs are C{\sc ii}-dominated for $A_V^{\rm tot}\lesssim1.5\,{\rm mag}$ (with $n_{\rm H}\sim600\,{\rm cm}^{-3}$ and $L\sim1.3\,{\rm pc}$), C{\sc i}-dominated for $1.5\lesssim A_V^{\rm tot}\lesssim3.2\,{\rm mag}$ (with $600\lesssim n_{\rm H}\lesssim 3700\,{\rm cm}^{-3}$ and $0.4\lesssim L\lesssim1.3\,{\rm pc}$) and CO-dominated for $A_V^{\rm tot}\gtrsim3.2\,{\rm mag}$ ($n_{\rm H}\gtrsim3700\,{\rm cm}^{-3}$ and $L\lesssim0.4\,{\rm pc}$). 

\begin{table*}
	\centering
	\caption{Fractional abundances of species as calculated for the `Reference ISM' defined in \S\ref{ssec:ref} adopting \avpdf\ and \avnh\ relations described in \S\ref{ssec:approach}. Columns $1{-}6$ refer to the shapes of {\avpdf}, mean ($M\equiv\overline{A_V}$), width ($\sigma$), median ($m$), mode ($\cal M$) and \avnh, respectively. Columns $7{-}11$ correspond to the average fractional abundances of species of CO, C{\sc i}, C{\sc ii}, H{\sc i} and H$_2$.
	}
	\label{tab:firstresults}
	\begin{tabular}{ccccccccccc}
		\hline
        \avpdf\ & $M$ & $\sigma$ & $m$ & ${\cal M}$ & \avnh\ &$f_{\rm CO}$ & $f_{\rm CI}$ & $f_{\rm CII}$ & $f_{\rm HI}$ & $f_{\rm H_2}$ \\ 
                 &&&  & &    & $\times10^{-5}$ & $\times10^{-5}$ & $\times10^{-5}$ & & \\ 
		\hline
        \multirow{3}{*}{\caseone} & \multirow{3}{*}{0.75} & \multirow{3}{*}{0.5} & \multirow{3}{*}{0.66} & \multirow{3}{*}{0.52} & Upper bound &0.0336 & 0.898 & 9.094 & 0.694 & 0.153 \\ 
        &&&&& Average & 0.0900 & 1.855 & 8.051 & 0.548 & 0.226 \\ 
        &&&&& Lower bound & 0.1569 & 2.616 & 7.210 & 0.461 & 0.269 \\[2ex] 
        \multirow{3}{*}{\casetwo} & \multirow{3}{*}{4.0} & \multirow{3}{*}{0.8} &\multirow{3}{*}{2.90} & \multirow{3}{*}{1.53} &  Upper bound &4.471 & 2.981 & 2.536 & 0.114 & 0.442 \\ 
        &&&& &Average& 5.317 & 2.842 & 1.827 & 0.080 & 0.460 \\ 
        &&&& &Lower bound& 5.800 & 2.714 & 1.470 & 0.065 & 0.467 \\ 
		\hline
	\end{tabular}
\end{table*}

\subsection{Convergence test}

The accuracy of the method is examined by performing a convergence test for the $q$ number of total PDR simulations needed to sample a given \avpdf. First, we lower this number by one half ($q=300$), implying a 0.02 dex logarithmic resolution in the $n_{\rm H}$ space. Second, we double the initial $q$ number ($q=1200$), corresponding to 0.005 dex resolution, at which point additional PDR simulations are run. 

In all cases, the changes in abundance stated in Table~\ref{tab:firstresults} (see \S\ref{sec:results}) between $q=300$ and $q=1200$ are always $\lesssim2\%$, and between $q=600$ and $q=1200$ are $\lesssim1\%$, indicating convergence. It is interesting to note that these changes are related with the different carbon phases, while variations in the H{\sc i} and H$_2$ abundances for different $q$ are negligible.

\section{Results}
\label{sec:results}

\subsection{Abundances for reference environmental parameters}
\label{sec:refavpdfs}

Table~\ref{tab:firstresults} lists the derived fractional abundances of species, $f_{\rm sp}$, for the two different {\avpdf}s considered here for the constant `environmental' parameters defined in \S\ref{ssec:ref}.
The \caseone\ \avpdf\ describes a diffuse, low density medium which is found to be rich in H{\sc i} and C{\sc ii}. When varying the \avnh\ relation within its bounds, the abundance of the more rare species CO, C{\sc i}, and H$_2$ are affected at a factor of $\lesssim 2{-}3$ level.
The \casetwo\ \avpdf\ describes a denser molecular cloud and is found to be rich in H$_2$ and contains a mix of C{\sc ii}, C{\sc i} and CO. We note that in all cases $f_\textrm{CO}+f_\textsc{Ci}+f_\textsc{Cii}\approx10^{-4}$, which is the carbon abundance adopted in the PDR simulations (Table~\ref{tab:abun}), and $f_\textsc{Hi}+2f_{\rm H_2}\approx1$. Within the boundary of the \avnh\ relation, the abundances of the rare species H{\sc i} and C{\sc ii} change at a factor $\lesssim 1.5$ level.

\begin{figure*}
\centering
\includegraphics[width=0.32\linewidth]{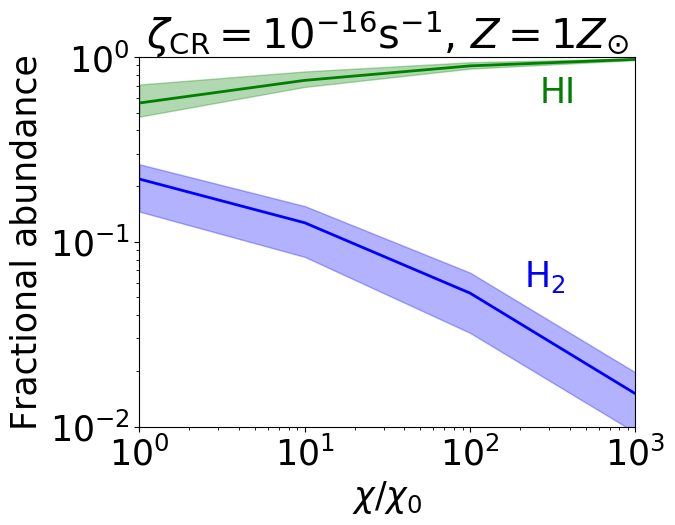}
\includegraphics[width=0.32\linewidth]{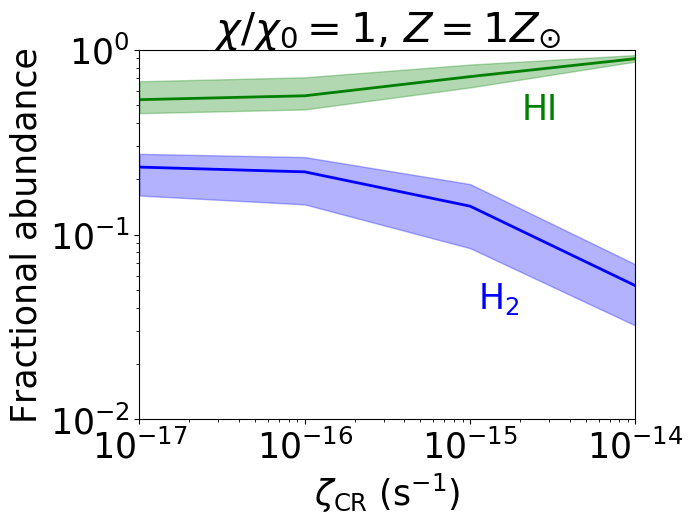}
\includegraphics[width=0.32\linewidth]{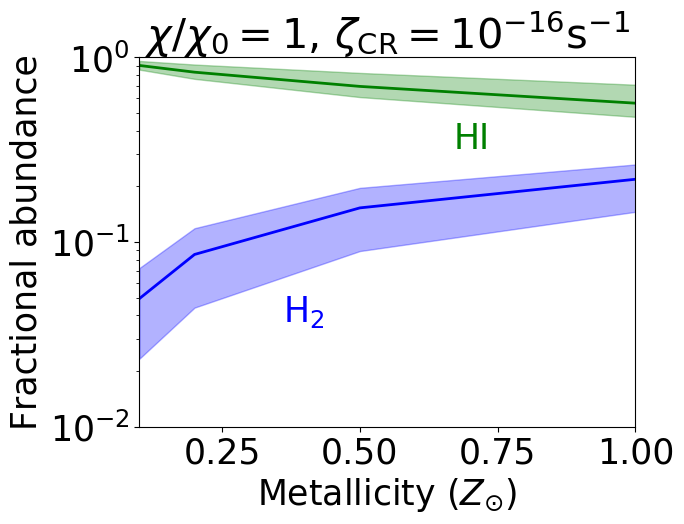}
\includegraphics[width=0.32\linewidth]{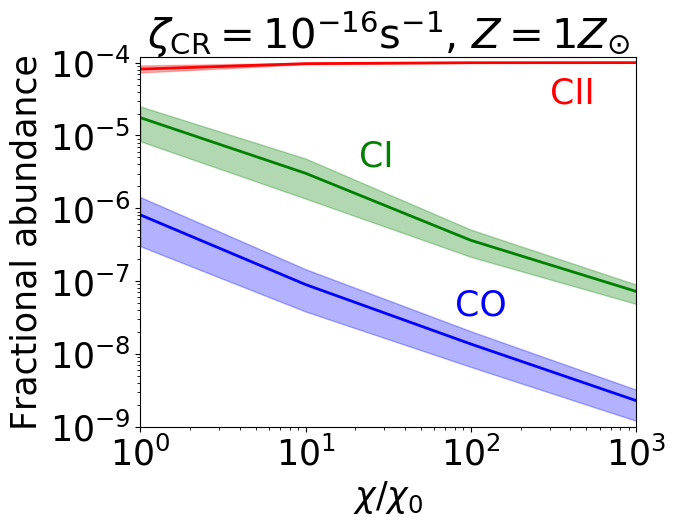}
\includegraphics[width=0.32\linewidth]{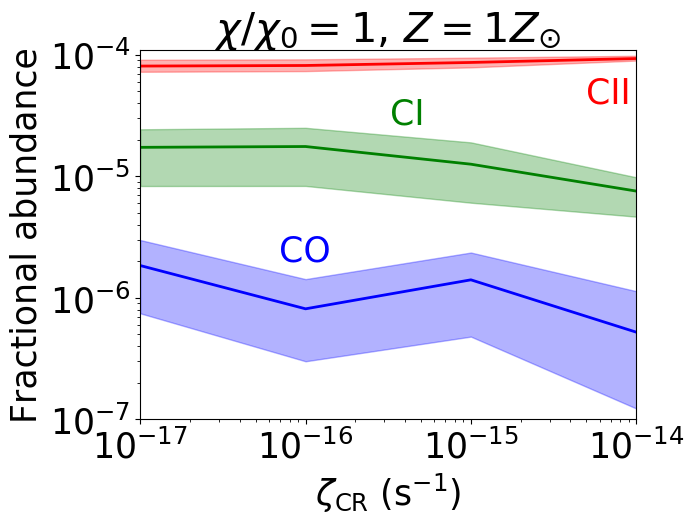}
\includegraphics[width=0.32\linewidth]{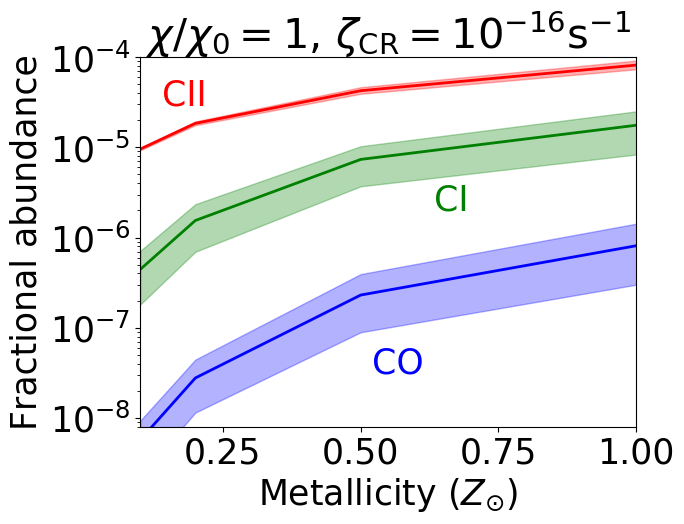}
\caption{Fractional abundances of H{\sc i} and H$_2$ (top row), and C{\sc ii}, C{\sc i}, CO (bottom row) for the \caseone\ \avpdf\ representing a diffuse, low density, atomic-dominated ISM as function of the FUV radiation field (left panels), cosmic-ray ionization rate (middle panels) and metallicity (right panels). The `Reference ISM' model considers $\zeta_{\rm CR}=10^{-16}\,{\rm s}^{-1}$, $\chi/\chi_0=1$ and $Z=1Z_{\odot}$. In each column we vary one of the free parameters while keep the other two fixed at their reference values.  Solid lines correspond to the average \avnh\ relation and the shaded region in each case is determined by the upper and lower bounds, as shown in Fig.~\ref{fig:avnh}. Under all conditions, the ISM is rich in H{\sc i} and C{\sc ii} due to the low $n_{\rm H}$ associated with \caseone\ \avpdf.}
\label{fig:resultscase1}
\end{figure*}

\begin{figure*}
\centering
\includegraphics[width=0.32\linewidth]{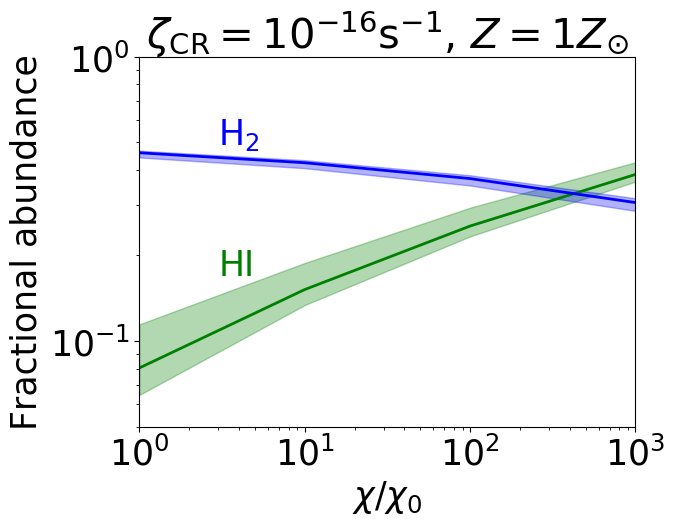}
\includegraphics[width=0.32\linewidth]{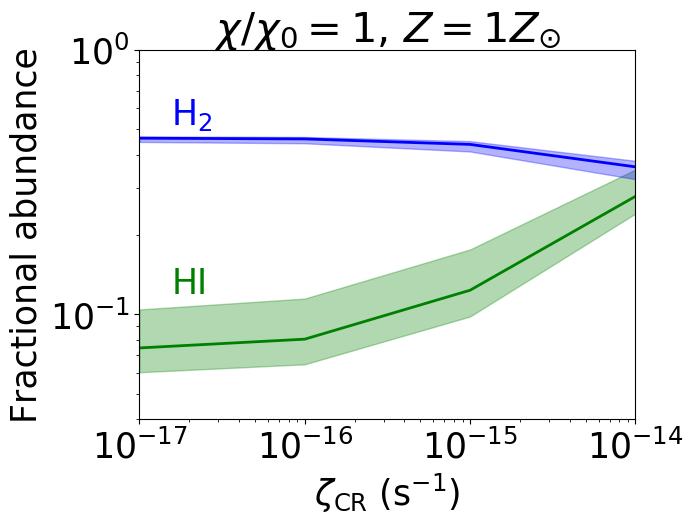}
\includegraphics[width=0.32\linewidth]{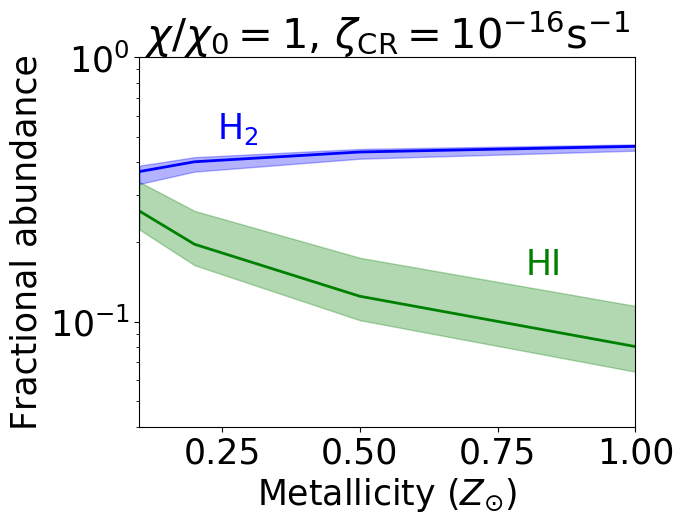}
\includegraphics[width=0.32\linewidth]{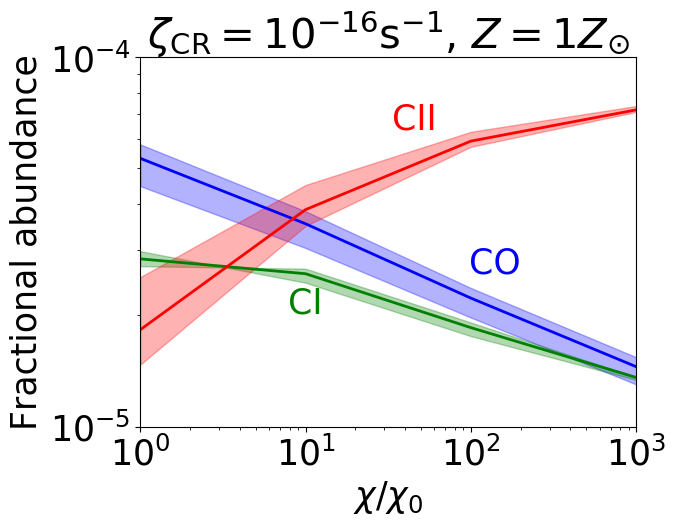}
\includegraphics[width=0.32\linewidth]{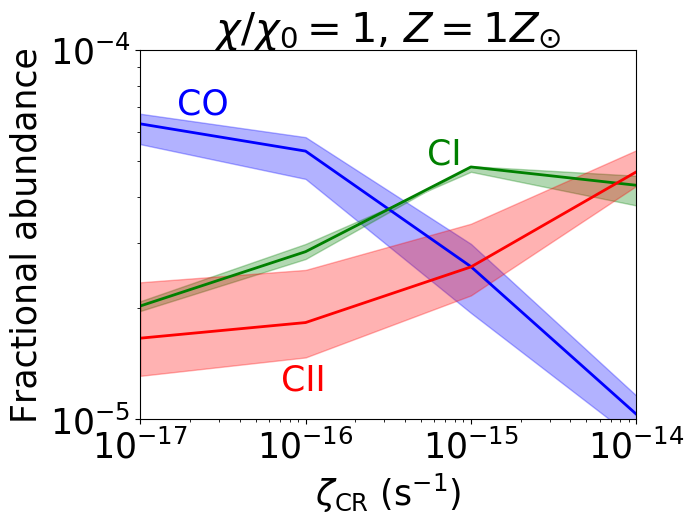}
\includegraphics[width=0.32\linewidth]{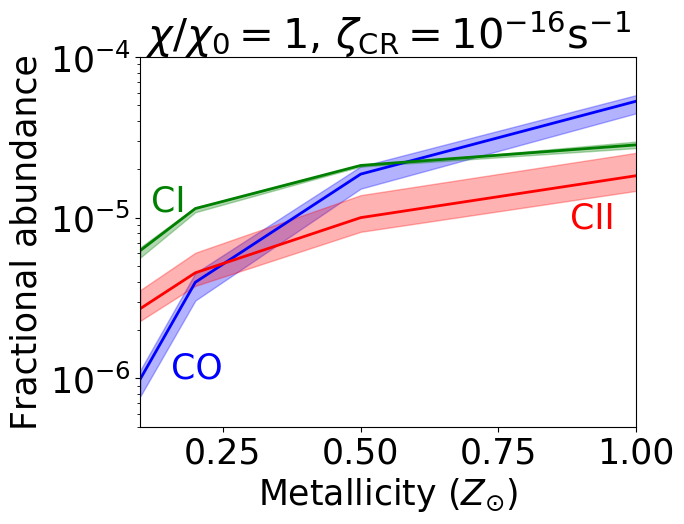}
\caption{Same as Fig.~\ref{fig:resultscase1} but now for the \casetwo\ \avpdf\ representing a denser molecular cloud. The cloud remains almost always molecular (i.e. H$_2$ dominates), however the carbon phase is more strongly dependent on the environmental conditions specified by the FUV intensity, the cosmic-ray ionization rate and the metallicity. This implies that existing methods of tracing H$_2$-rich gas may require adjustments depending on environment.}
\label{fig:resultscase2}
\end{figure*}

With this `reference' as a starting point, three suites of PDR calculations are performed in which the incident FUV radiation field $\chi$, the cosmic ray ionization rate $\zeta_{\rm CR}$, and the metallicity $Z$ are varied as free parameters. In particular, we consider four different intensities of the incident FUV radiation field ($\chi/\chi_0=10^0-10^3$), four different cosmic-ray ionization rates ($\zeta_{\rm CR}=10^{-17}-10^{-14}\,{\rm s}^{-1}$), and four different metallicities ($Z/Z_{\odot}=0.1,\,0.2,\,0.5,\,1$). Figure~\ref{fig:resultscase1} shows the results for the \caseone\ \avpdf\ and Figure~\ref{fig:resultscase2} those for \casetwo.

\subsection{Varying the FUV radiation field intensity}
\label{ssec:fuv}

In the first suite of calculations, the external FUV radiation field $\chi$ is varied as free parameter while $\zeta_{\rm CR}=10^{-16}\,{\rm s}^{-1}$ and $Z=1\,Z_{\odot}$ are kept constant.  The left columns of Figs.~\ref{fig:resultscase1} and \ref{fig:resultscase2} show the average abundance of H{\sc i}, H$_2$, C{\sc ii}, C{\sc i} and CO as a function of $\chi/\chi_0$ for the \caseone\ and \casetwo\ PDF models, respectively. Since the \caseone\ PDF corresponds to a diffuse, low density medium, it is expected to be predominantly in atomic form with high abundances of C{\sc ii}.   Indeed, as can be seen in Fig.~\ref{fig:resultscase1}, for $\chi/\chi_0=1$, $f_\textsc{Hi}$ is already $\gtrsim0.5$ and $f_\textsc{Cii} \gtrsim 8\times10^{-5}$ (which is $\sim80\%$ of carbon) for all \avnh\ relations. The abundances of H$_2$, C{\sc i} and CO differ by a factor between two to four as a result of the adopted upper and lower \avnh\ relations for the low $A_V$ regime. However, the abundances of these species are already much smaller when compared to the more dominant H{\sc i} and C{\sc ii}, respectively, and such a difference can therefore be considered negligible. H$_2$, C{\sc i} and CO decrease monotonically with increasing FUV radiation field intensity. For the case of $\chi/\chi_0=10^3$, this low density medium is H{\sc i}- and C{\sc ii}-dominated by $\gtrsim99\%$.

The much denser molecular cloud described by the \casetwo\ PDF (left column of Fig.~\ref{fig:resultscase2}) is dominated by H$_2$ for FUV radiation fields up to $\chi/\chi_0\approx300$, above which it becomes H{\sc i}-dominated. For $\chi/\chi_0\lesssim7$ the cloud is dominated by CO, above which is dominated by C{\sc ii}. It is interesting to note that the cloud never becomes C{\sc i}-dominated as the FUV field increases. Instead, it is predicted to be dominated by warm gas rich in H$_2$ and C{\sc ii}. The largest fraction of C{\sc i} is $\sim2.8\times10^{-5}$ and for $\chi/\chi_0=1$.  Due to the overall high density, the shaded region due to the decreasing \avnh\ bounds is thinner than in {\caseone} described above.

For the \caseone\ PDF, the average abundances of H$_2$ and of the carbon phases change considerably with varying the radiation field. This is because the low density medium is translucent at low values of $A_{\rm V}$, and it implies drastic changes in the abundance ratios of CO/H$_2$, C{\sc i}/H$_2$ and C{\sc ii}/H$_2$. For the \casetwo\ PDF, none of the abundances experience such large changes and the abundance of H$_2$ scales closely with CO and C{\sc i} (but not C{\sc ii}). 

\subsection{Varying the cosmic-ray ionization rate}

As a next application, we study the change in abundance of key species as a function of $\zeta_{\rm CR}$. The middle columns of Figs.~\ref{fig:resultscase1} and \ref{fig:resultscase2} show the results of these calculations. For the \caseone\ PDF, the abundance of C{\sc ii} remains very high regardless of $\zeta_{\rm CR}$. Likewise, H{\sc i} increases and this makes the low density medium to be $\gtrsim70\%$ in the atomic phase, particularly for $\zeta_{\rm CR}\gtrsim10^{-15}\,{\rm s}^{-1}$. For $\zeta_{\rm CR}\approx10^{-15}\,{\rm s}^{-1}$, CO has a local maximum (although its abundance is overall very low), which is most likely attributed to the formation of CO via the OH channel as described in \citet{Bisb17}. It is interesting to note that C{\sc i} remains nearly constant as $\zeta_{\rm CR}$ increases and that C{\sc i} is always one to two orders of magnitude more abundant than CO.

In contrast, the denser molecular cloud described by the \casetwo\ PDF remains predominantly molecular for all $\zeta_{\rm CR}$. The carbon phase sensitively depends on the cosmic-ray ionization rate. Above $\zeta_{\rm CR}\gtrsim3\times10^{-16}\,{\rm s}^{-1}$, CO is effectively destroyed and carbon is either in C{\sc i} or C{\sc ii}. For $3\times10^{-16}<\zeta_{\rm CR}<5\times10^{-15}\,{\rm s}^{-1}$ the carbon is in C{\sc i}, while for higher $\zeta_{\rm CR}$ it is in C{\sc ii}. As already noted by \citet{Bisb15}, for moderately enhanced cosmic-ray ionization rates, C{\sc i} becomes a better tracer of H$_2$-rich gas than CO. In this regime, C{\sc ii} is also very abundant, however it may well originate from other heating sources such as FUV radiation (see \S\ref{ssec:fuv}), shocks or X-rays \citep{Mack18} and, therefore, is not always connected to the presence of H$_2$.

\subsection{Varying the metallicity}

As a third test, we investigate how the abundances of species change with lower metallicity ($Z$), while keeping FUV intensity and the cosmic-ray ionization rate fixed. In this work, we take the depths of clouds at low metallicity to be the same as those for solar metallicity for a given density.

The right columns of Figs.~\ref{fig:resultscase1} and \ref{fig:resultscase2} show the corresponding calculations. The low density medium represented by the \caseone\ PDF is always found to be atomic dominated and as $Z$ decreases the abundance of H{\sc i} increases. Almost all carbon is in form of C{\sc ii}. As metallicity decreases, the abundances of all carbon phases also decrease in parallel and no transition from one phase to another is observed. At all metallicities, the fractional abundance of CO relative to hydrogen is very small ($\lesssim5\times10^{-7}$).

The denser molecular cloud (\casetwo) remains fully molecular for all metallicities. At solar metallicity, the carbon is two to three times more in CO, than in C{\sc i} or C{\sc ii}. For $Z<0.5\,Z_{\odot}$ it is dominated by C{\sc i}. For $Z\sim0.1\,Z_{\odot}$, C{\sc i} has two times the abundance of C{\sc ii} and six times the abundance of CO. This is because H$_2$ self-shields from FUV radiation, whereas CO is shielded by dust, which is reduced in low metallicity environments. In addition, the column for CO self-shielding builds up deeper into cloud due to the overall lower abundances of C and O. Therefore, CO photodissociates creating a surplus of C{\sc i} and C{\sc ii} \citep{Malo88,Pak98,Bola99}.

\subsection{Considering an \avpdf\ similar to Taurus}
\label{ssec:case3}

\begin{figure*}
\centering
\includegraphics[width=\linewidth]{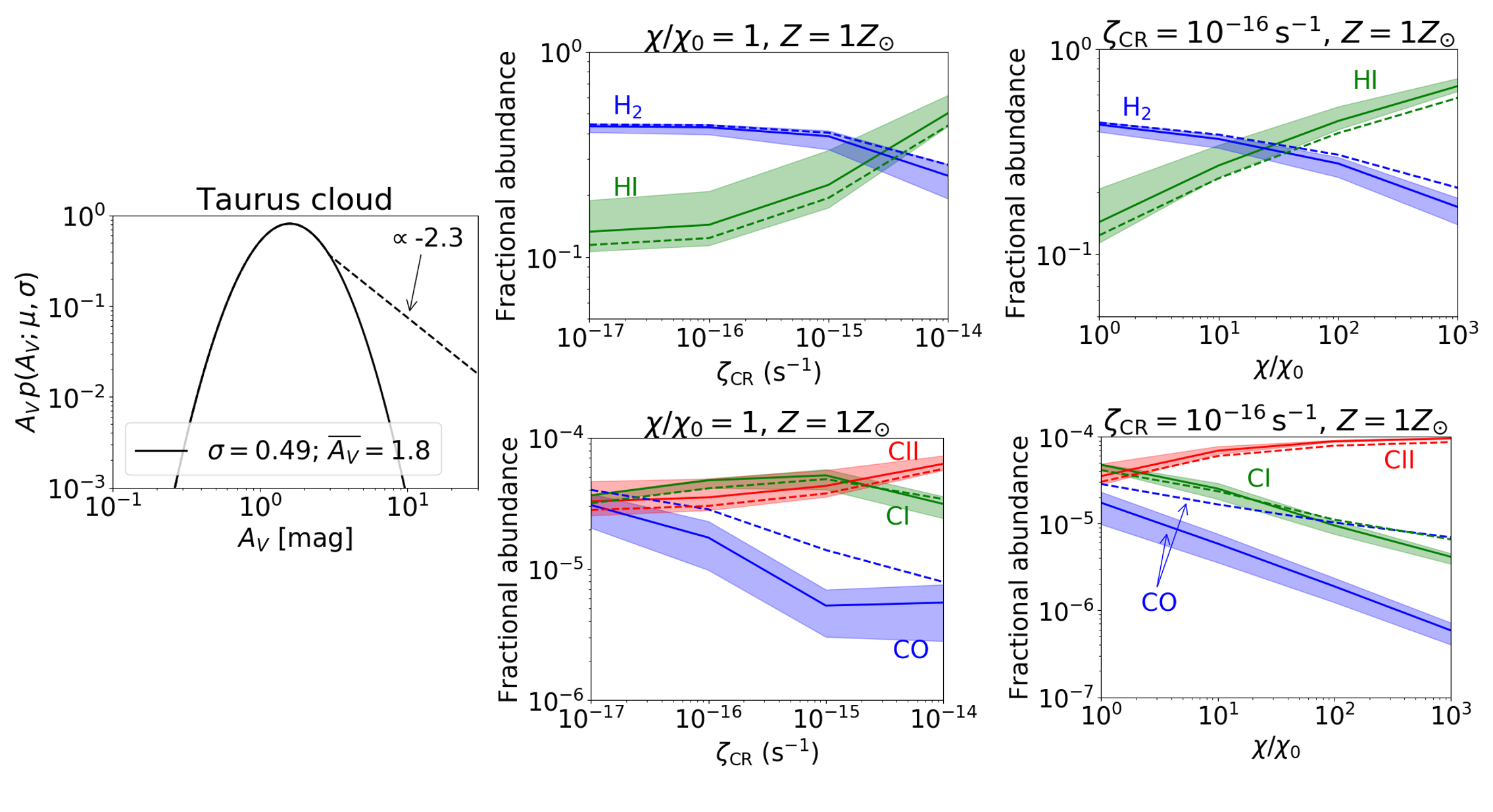}
\caption{\emph{Left panel:} \avpdf\ with $\overline{A_{\rm V}}=1.8\,{\rm mag}$ and $\sigma=0.49$ representative of the Taurus star-forming cloud \citep{Kain09}. Dashed line shows the power-law tail which has a slope $\propto-2.3$ (in the units of $dp/dA_V$) for $A_{V}>3\,{\rm mag}$ \emph{Right panels:} Similarly to Figs.~\ref{fig:resultscase1} and \ref{fig:resultscase2}, these four panels show how the average abundances vary as a function of $\zeta_{\rm CR}$ (middle column) and UV radiation (right column) for H{\sc i} and H$_2$ (top row) and the carbon cycle (bottom row). The metallicity is kept fixed at solar. Solid lines and their associated shadowed regions correspond to the log-normal component only. The dashed line corresponds to the \avpdf\ when the power-law tail is considered and for which the `average' \avnh\ relation was used, to avoid confusion.}
\label{fig:resultscase3}
\end{figure*}

Lastly, we consider an \avpdf\ with $\overline{A_{\rm V}}=1.8\,{\rm mag}$ and width $\sigma=0.49$ which parametrises the column density distribution of the nearby Taurus star-forming cloud as observed by \citet{Kain09} at $\sim0.1\,{\rm pc}$ resolution. For our `average' \avnh\ relation, this corresponds to a density of $n_{\rm H}\sim300\,{\rm cm}^{-3}$ and a size of $L\sim2\,{\rm pc}$.  Furthermore, we consider the impact of the power-law tail that has been observed for this region \citep{Kain09} on the overall chemical composition. The power-law tail is shown in the left panel of Fig.~\ref{fig:resultscase3} (dashed line) and has a slope of about $-2.3$ (in the units of $dp/dA_V$) at $A_V>3\,{\rm mag}$.
While we do not attempt to examine in detail the physical conditions prevailing in the ISM of Taurus, it is interesting to explore how the average abundances for an observed \avpdf\ change as a function of the free parameters considered in this work. 

The middle and right columns of Fig.~\ref{fig:resultscase3} show how the abundances scale with varying radiation field or cosmic ray ionization rates while keeping the metallicity fixed at solar. Solid lines along with their associated shadowed regions correspond to the log-normal component only without the contribution of the power-law tail. Dashed lines show results of the entire \avpdf\ when the power-law tail has been considered (for which the `average' \avnh\ relation has been taken into account only, to avoid confusion). From the \avpdf\ it is seen that this cloud remains dominated by H$_2$ for moderate-to-high values of cosmic-ray ionization rates ($\zeta_{\rm CR}\lesssim3\times10^{-15}\,{\rm s}^{-1}$), but is CO-poor particularly for $\zeta_{\rm CR}>10^{-16}\,{\rm s}^{-1}$ (Fig.~\ref{fig:resultscase3} middle column).

Cosmic-rays attenuate as a function of column density which has consequences in the CO/H$_2$ abundance ratio.
For example, \citet{Pado18} find that the cosmic-ray ionization rate for $A_V>4\,{\rm mag}$ may be $\zeta_{\rm CR}\lesssim10^{-17}\,{\rm s}^{-1}$. Thus, the densest part of the Taurus star-forming cloud may have lower $\zeta_{\rm CR}$ values than those examined here. This in turn implies higher CO abundances than those predicted in Fig.~\ref{fig:resultscase3}. It is interesting to note that the molecular gas is equally rich in C{\sc i} and C{\sc ii} until it transits to the atomic phase, where it becomes C{\sc ii}-rich. When the power-law component is taken into consideration, it is found that the abundance of H$_2$ increases by a small amount which is reflected by the small decrease in H{\sc i} abundance. The most striking feature, however, is the sudden increase in CO abundance for all $\zeta_{\rm CR}$ and $\chi/\chi_0$ values explored. In particular, for $\zeta_{\rm CR}\lesssim3\times10^{-17}\,{\rm s}^{-1}$ it is found that the molecular gas is CO-rich contrary to the findings of the log-normal component-only. 

The cloud remains also H$_2$-rich for FUV intensities up to  $\chi/\chi_0\sim10-40$, as shown in the right column of Fig.~\ref{fig:resultscase3}. For values of $\chi/\chi_0\sim1$, C{\sc i} and C{\sc ii} (and CO when the power-law tail is considered) have comparable abundances but for higher intensities it becomes (and stays) very C{\sc ii}-rich. There is thus an extended range of $\zeta_{\rm CR}$ ($\sim10^{-17}-10^{-15}\,{\rm s}^{-1}$) and $\chi/\chi_0$ ($\sim1-10$) in which the abundance ratio of C{\sc ii}/H$_2$ is high. When compared to the \casetwo\ {\avpdf} (which has $\sim2.2$ times higher $\overline{A_V}$ and $\sim1.6$ times larger $\sigma$) C{\sc ii}-dominated H$_2$-rich gas is observed for a more extended range of $\chi/\chi_0$ ($\sim10-300$), although the molecular gas is either CO-, C{\sc i} or C{\sc ii}-dominated depending on the value of $\zeta_{\rm CR}$.

\section{Difference to uniform density calculations}
\label{sec:validation}


In this section, we investigate whether the abundances of lognormal {\avpdf}s can be adequately estimated by single uniform density PDR calculations as compared to the methodology presented in \S\ref{ssec:approach} of models that take into account the full distribution of extinction columns and associated volume densities. In particular, we consider lognormal {\avpdf}s and determine the chemical abundances at the mean, median and mode extinction columns (which are defined by Eqns.\ \mbox{(\ref{eq:var1})--(\ref{eq:var3}))}. We keep the median constant ($m=3$ mag) and consider six different widths ($\sigma=0.01, 0.1, 0.2, 0.5, 1.0, 1.2$), which also implies variations in the mean ($\overline{A_V}$). We thus construct six different PDF functions, namely $p_1-p_6$ corresponding to each different width, respectively. In the special case where $\sigma$ tends to zero, the PDF tends to a $\delta$-function and thus the aforementioned quantities become identical\footnote{In symmetrical distributions, such as a Gaussian one, the mean, median and mode are always equivalent.}.

\begin{figure*}
\centering
\includegraphics[width=0.32\linewidth]{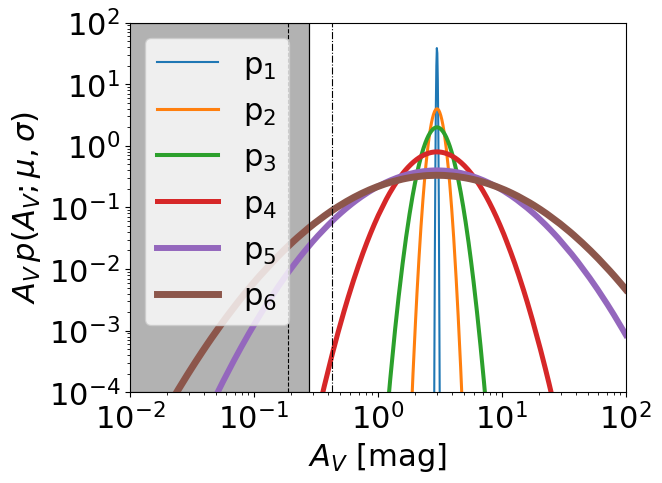}
\includegraphics[width=0.32\linewidth]{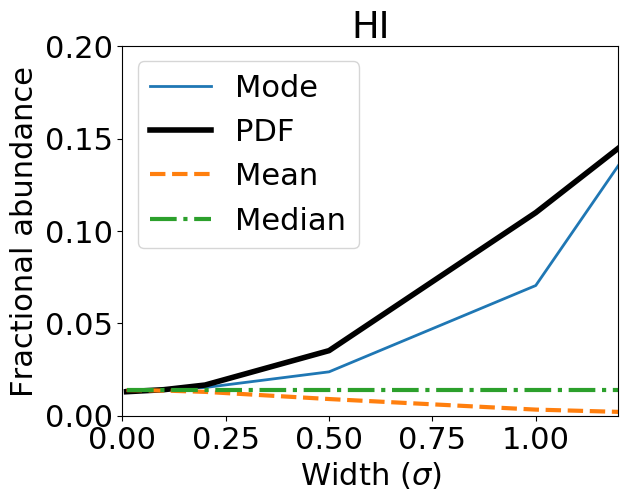}
\includegraphics[width=0.32\linewidth]{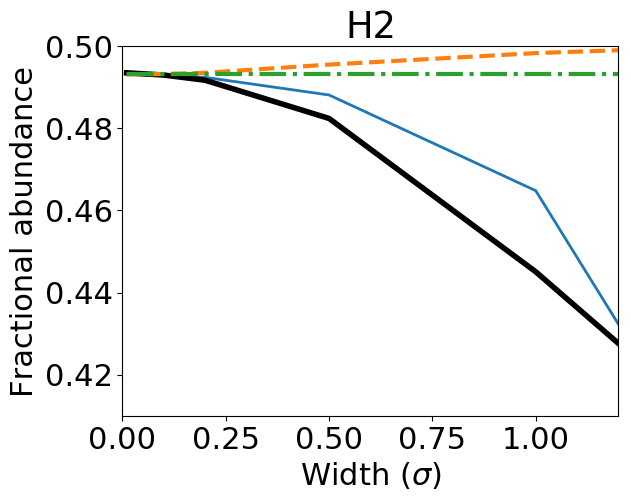}
\includegraphics[width=0.32\linewidth]{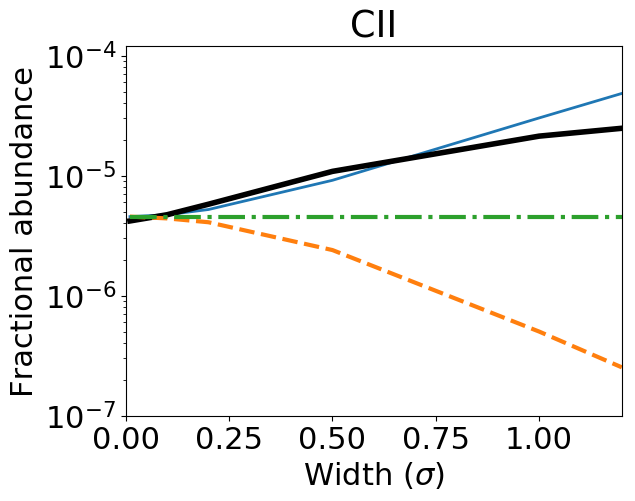}
\includegraphics[width=0.32\linewidth]{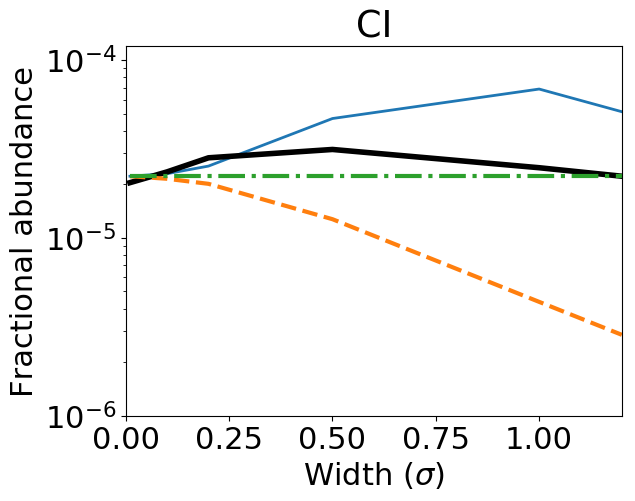}
\includegraphics[width=0.32\linewidth]{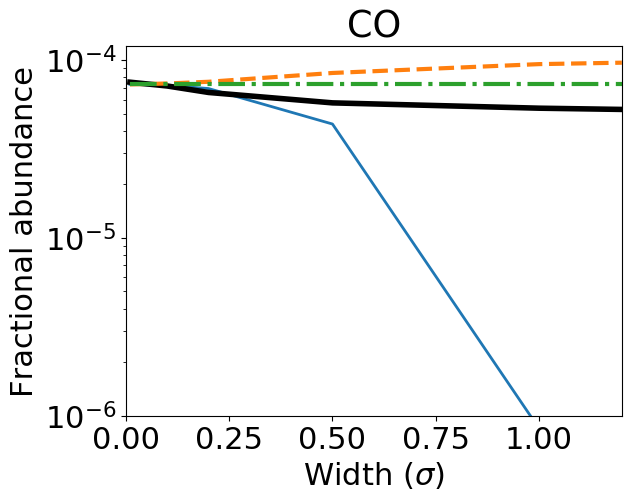}
\caption{\emph{Top left:} \avpdf\ with fixed median of $m=3$ mag and varying widths $\sigma$, which implies variations in the mean $\overline{A_V}$. \emph{Top middle and right panels} show the abundances of H{\sc i} and H$_2$ respectively as the width increases while keeping the same mode, and for the four different approaches discussed in \S\ref{sec:validation}. \emph{Bottom left, middle and right panels} show the abundances of C{\sc ii}, C{\sc i} and CO respectively. As expected, for small $\sigma$ all different approaches converge.}
\label{fig:validation}
\end{figure*}

The top left panel of Fig.~\ref{fig:validation} shows the shapes of the considered {\avpdf}s. For each $\sigma$, the abundances of H{\sc i}, H$_2$, C{\sc ii}, C{\sc i} and CO are calculated using uniform density PDRs corresponding to the mean, median and mode, as well as using the \S\ref{ssec:approach} method (referred to as `PDF' in Fig.~\ref{fig:validation}). All these results are shown in the remaining panels of Fig.~\ref{fig:validation}. For the distributions with $m=3$ mag, values of $\sigma\lesssim0.3$ show no appreciable difference in the abundances of species as calculated using a particular method.

As $\sigma$ increases, the difference between the derived abundances based on the above methods becomes significant. It is interesting to note that the abundances calculated with uniform density PDRs based on the values of the mean and the median always overestimate the abundances of H$_2$ and CO, whereas they always underestimate the abundances of atomic species (H{\sc i}, C{\sc i}) and the abundance of C{\sc ii}. Although the mode and `PDF' approaches follow the same trend for H{\sc i}, H$_2$ and C{\sc ii} with increasing $\sigma$, there is a discrepancy when calculating the abundances of C{\sc i} and CO. This is because the column density of this species peaks only for a particular range of $A_V$ (see \S\ref{sec:ccycle}) and it is thus more strongly dependent on how this range of $A_V$ compares to the $A_V$ corresponding to the aforementioned approaches. Overall, the mode $A_V$ provides the best approach for single PDR calculations, but significant differences with the full PDF recommending using the here presented method that takes full account of the \avpdf.

\section{Discussion}
\label{sec:discussion}

The method presented is quick and robust. Since it considers an \avpdf\ distribution as input, it is able to give a more realistic estimation of the average abundances of species compared to uniform (1D) PDR calculations with densities representing the average number density of a particular object. Thus, it can be used to provide fast and reasonable estimations of ISM structures spanning from a few parsec to kiloparsec scales. However, there are several assumptions and limitations made at this stage, which can be improved in future works.

In the most basic framework adopted here, it has been assumed that the chemical evolution time, $t_{\rm chem}$, is significantly smaller than the hydrodynamical evolution time, $t_{\rm hydro}$. The condition of $t_{\rm chem}\ll t_{\rm hydro}$ corresponds to quiescent (non-star-forming) molecular clouds the ISM of which is considered to be in statistical equilibrium. However, it is known that turbulence, which is always present in real clouds, can affect the chemical evolution of the ISM by cycling material from low to high density gas and back, and thereby altering the abundances of species. The H$_2$ formation time is $t_{\rm H_2}\!\sim\!10^9\,{\rm yr}/(n_{\rm H}/{\rm cm}^{-3})$ \citep{Holl79} which for a gas cloud of density $n_{\rm H}\sim100\,{\rm cm}^{-3}$ gives $t_{\rm H_2}\sim10^7\,{\rm yr}$. This time is needed for the atomic-to-molecular transition to reach equilibrium \citep{Holl99}. On the other hand, for quiescent molecular clouds, the turbulent diffusion time (over which turbulence has significantly decayed) is a few $10^6\,{\rm yr}$ \citep{Xie95} and thus comparable to the H$_2$ formation time for this specific density. Thus in a turbulent medium, equilibrium may be reached at different times, as discussed below. 

Modern 3D numerical simulations of turbulent clouds that include time dependent chemistry find that turbulence can speed up the H$_2$ formation time at a respective $n_{\rm H}$ when compared to a quiescent case \citep{Glov07a,Glov07b,Glov10}. Recent simulations of Milky Way-type molecular clouds by \citet{Seif17} indicate that such clouds may contain molecular hydrogen in regions with number densities of $n_{\rm H}\sim30\,{\rm cm}^{-3}$ as a result of turbulent mixing, despite the fact that the local H$_2$ formation time there is $t_{\rm H_2}\sim10^8\,{\rm yr}$. This may have secondary effects in the connection of {\avpdf}s with the average abundances of species as presented here, since our PDR calculations do not account for such turbulent mixing and are thus not suitable to study turbulent media. For example, \citet{Bial17} find that once turbulence becomes supersonic, the development of strong density fluctuations distort the atomic-to-molecular transition. They further find that the mean and the median of the H{\sc i} column densities are affected by supersonic turbulence, but since turbulence decays more quickly as density increases, any changes to the \avnh\ relation should predominantly affect low values of $A_{\rm V}$. Therefore, modifications of our methodology may be required only at the low $A_V$ end to account for turbulent mixing in regions of supersonic turbulence.

In all applications discussed in \S\ref{sec:results}, the input parameters ($\chi$, $\zeta_{\rm CR}$, $Z$) are assumed to be uniform. This condition may hold only for small ISM patches and individual molecular clouds, unless there are strong, embedded UV sources. Despite this caveat, uniform boundary conditions are frequently adopted in 3D simulations of quiescent or star-forming clouds \citep[e.g.][]{Glov10, Clar13, Wu17, Bisb17, Bisb17b, Bisb18}. Likewise, the cosmic-ray ionization rate is considered constant with no attenuation as a function of $A_V$ \citep[e.g.][]{Clar13, Bisb17}, and metallicity is assumed constant \citep[e.g.][]{Glov12, Papa18}. However, when studying the ISM at larger (kpc) scale, the above assumptions may not hold as all three aforementioned input parameters may vary spatially. For example, \citet{Wolf03} provide an analytical model of how the FUV radiation field, cosmic ray ionization rate and metallicity vary with galactic radius in the Milky Way. Their model shows that all these quantities vary significantly (e.g. up to 20 times) in different places of the Galaxy and therefore should be accounted for when studying large ISM patches. Recent numerical simulations by \citet{Smit14}, \citet{Giri16} and \citet{Li18} also highlight the complexity of the ISM in this regard. Thus it can be expected that the `environmental' parameters considered in our framework, need not always be assumed as constant values, but rather vary according to some distribution as well, which should be combined with the respective {\avpdf}. In this situation, Eqn.~(\ref{eqn:fsp}) should be modified as follows:
\begin{eqnarray}
\label{eqn:fsp2}
\langle f_{\rm sp}\rangle=\frac{\sum_j^\ell f_{{\rm sp},j} \cdot{\rm PDF}_j}{\sum_j^{\ell}{\rm PDF}_j},
\end{eqnarray}
where PDF$_j$ is the probability distribution function of the `environmental' parameters ($\chi$, $\zeta_{\rm CR}$, $Z$) and $\ell$ is the number of different PDR suites, each suite consisting of $q$ simulations for a particular input parameter.

An important advancement in the present method will be the inclusion of radiative transfer to calculate the line emission of key tracers such as [CII], [CI], different CO transitions and others that are captured by observational studies. Several algorithms are now moving into this direction, as it provides the most direct way of comparing observational data with numeric simulations. Such tools include {\sc radex} \citep{vdTa07}, {\sc lime} \citep{Brin10}, {\sc radmc-3d} \citep{Dull12} and more recently {\sc PyRaTe} \citep{Trit18}. However, calculation of abundances using PDR chemistry resulting in more accurate level populations and therefore optical depths, is not explicitly done in the aforementioned codes. A future aim of this work is to couple the presented method with a radiative transfer scheme. This will allow the determination of average line intensities for complex ISM distributions and predict how fundamental quantities---such as CO or C{\sc i} conversion factors---change as function of various `environmental' parameters \citep[e.g.][]{Papa04,Bola13}. For example, the middle panels of Fig.~(\ref{fig:resultscase2}) show that as $\zeta_{\rm CR}$ increases, the carbon phase is more sensitive to the increment of $\zeta_{\rm CR}$ than the atomic-to-molecular transition is. As already highlighted by \citet{Bisb15, Bisb17}, this is a strong indicator that the common technique of using CO as a tracer of H$_2$ might be biased. In particular in high-redshift galaxies, that appear to have elevated $\zeta_{\rm CR}$ values, additional and more detailed examination is needed to establish the corresponding values of CO- and C{\sc i}-to-H$_2$ conversion factors. At the same time, observations of low metallicity environments, where CO emission is traditionally found to be very weak \citep[e.g.][]{Lero05,Schr12,Schr17}, would greatly benefit from detailed knowledge of how \mbox{(non-)} detected line intensities relate to ISM properties. 

\section{Conclusions}
\label{sec:conclusions}

We present a new numerical algorithm that uses \avpdf\ and \avnh\ relations as inputs to estimate the average abundances of species with look-up tables from a suite of pre-calculated 1D thermo-chemical photodissociation region (PDR) calculations. This approach is much faster to full hydrochemical calculations where the computational cost is very high. It is thus a quick and alternative tool to estimate the average conditions of large ISM scales and assess trends with parameters, suitable for extragalactic studies.

Two hypothetical, lognormal \avpdf\ distributions are examined: a first one corresponding to a low density, atomic-dominated medium (with $\overline{A_V} = 0.75$ mag and width $\sigma = 0.5$) and a second one corresponding to a dense molecular cloud (with $\overline{A_V} = 4$ mag and $\sigma = 0.8$). For each {\avpdf}, the impact of different FUV radiation fields, cosmic-ray ionization rates and metallicities on the abundance of species was investigated. A third distribution (with $\overline{A_V} = 1.8$ mag and $\sigma = 0.49$) was considered describing the observed \avpdf\ of the Taurus molecular cloud.

In the case of the low density medium, the gas remains fully H{\sc i}- and C{\sc ii}-dominated. However, in the case of the denser molecular cloud, we find that, although the cloud remains entirely molecular, its carbon phase can either be CO, C{\sc i} or even C{\sc ii}-dominated, depending on the radiation field, cosmic ray rates and metallicity. For the particular case of the Taurus-like cloud {and when considering the log-normal component only,} we find that its molecular phase is equally C{\sc i} and C{\sc ii} for a wide range of cosmic-ray ionization rates at a range of low UV intensities. It is C{\sc ii}-rich for UV intensities up to a few for low cosmic-ray ionization rates, after which the medium becomes atomic. When considering the power-law tail in addition to the log-normal component, we find a small increase in H$_2$ but a more striking increase in the CO abundances. The here presented algorithm is flexible and computationally efficient, and thus can be a powerful tool to determine the chemical composition of large ISM regions without the necessity to perform detailed and expensive astrochemical calculations. 

We note that the algorithm may need to be modified to model kiloparsec-scale ISM regions. One reason for this is that supersonic turbulence can dynamically mix the gas so that a non-equilibrium approach is required. In addition, the input parameters of the external FUV radiation field, cosmic-ray ionization rate and metallicity are considered to be uniform for now. While this assumption seems reasonable for small ISM patches, it becomes challenged when considering the ISM at large scales. In principle, large scale variations can be accounted for by assuming distribution functions of the ISM properties and external parameters, and applying our algorithm ensembles of smaller ISM patches. Finally, coupling the present algorithm with radiative transfer will provide a powerful tool for studying the average intensities of emission lines of large ISM regions with complex substructure. This may be of particular importance for studies of high-redshift galaxies that often remain marginally resolved by observations and one-zone PDR calculations may fail to deliver reliably characterizations of their physical properties.

\section*{Acknowledgements}

The authors thank the referee for the useful comments which improved the clarity of this work. This work is supported by a Royal Netherlands Academy of Arts and Sciences (KNAW) professor prize, and by the Netherlands Research School for Astronomy (NOVA). TGB acknowledges funding by the German Science Foundation (DFG) via the Collaborative Research Center SFB 956 ``Conditions and impact of star formation''. The authors thank Alyssa Goodman, Desika Narayanan, Blakesley Burkhart, Shmuel Bialy, Amiel Sternberg, Daniel Seifried, Sebastian Haid, Annika Franeck, Stefanie Walch, Nicola Schneider, Volker Ossenkopf and Markus R\"ollig for insightful discussions. This research has made use of NASA's Astrophysics Data System.



\appendix

\section{The H{\sc i}-to-H$_2$ transition}
\label{sec:app}

The network of chemical reactions following the inclusion of the suprathermal formation of CH$^+$, described in \ref{ssec:supr}, destroys the H$_2$ molecule at low column densities, thereby shifting the atomic-to-molecular transition to higher $A_V$. To investigate this further, an isothermal PDR simulation of a uniform density cloud is performed ($T_{\rm gas}=50\,{\rm K}$, $n_{\rm H}=10^3\,{\rm cm}^{-3}$, $\chi/\chi_0=1$, $\zeta_{\rm CR}=10^{-16}\,{\rm s}^{-1}$) where the `suprathermal' routine is switched ON and OFF. 

\begin{figure}
    \centering
    \includegraphics[width=0.95\linewidth]{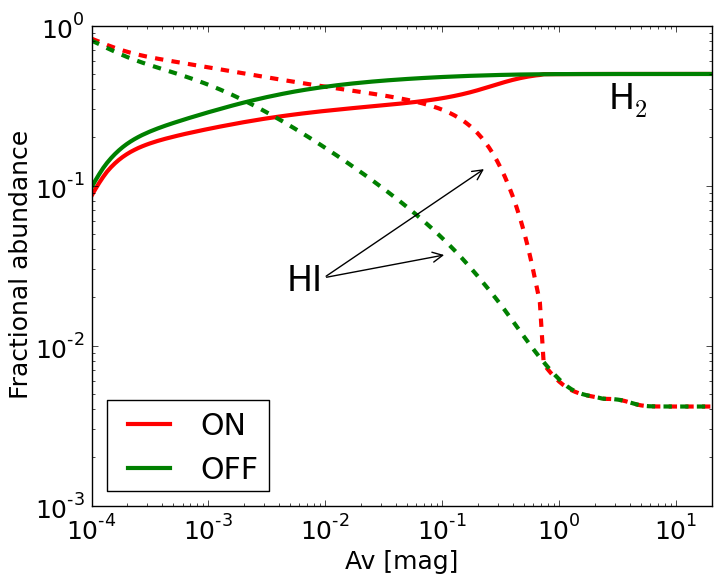}
    \includegraphics[width=0.98\linewidth]{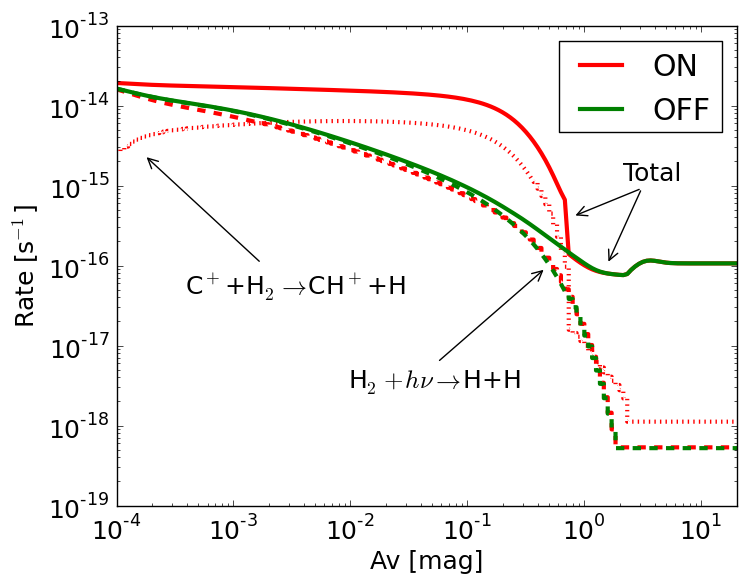}
    \caption{{\it Top:} Fractional abundances of atomic and molecular hydrogen as a function of $A_V$ for the isothermal PDR simulation where we switch ON (red lines) and OFF (green lines) the routine of suprathermal formation of CH$^+$. Solid lines show H$_2$ whereas dashed lines show H{\sc i} abundances. The H{\sc i}-to-H$_2$ transition occurs at $A_V\sim2\times10^{-3}\,{\rm mag}$ when the `suprathermal' routine is OFF, whereas it occurs at $A_V\sim6\times10^{-2}\,{\rm mag}$ when it is ON. {\it Bottom:} Destruction rates of H$_2$ as a function of $A_V$. Once the suprathermal routine is ON, H$_2$ is being destroyed in the $2\times10^{-3}\lesssim A_V\lesssim0.7\,{\rm mag}$ range via its reaction with C$^+$ (dotted line; see \S\ref{ssec:supr}), which dominates over the destructive reaction due to photodissociation (dashed lines). This shifts the H{\sc i}-to-H$_2$ transition at higher column densities as shown in the top panel.}
    \label{fig:app}
\end{figure}

Figure~\ref{fig:app} shows the results of the fractional abundances (top) and the corresponding destruction rates (bottom) of the H$_2$ molecule as a function of $A_V$. Once the suprathermal routine is switched ON, H$_2$ is destroyed by Reaction~\ref{reac:sup} (see discussion in \S\ref{ssec:supr}) which dominates over the H$_2$ photodissociation reaction, therefore delaying the formation of H$_2$-rich gas.

\section{The \avnh\ relation}
\label{sec:app2}

\begin{figure}
    \centering
    \includegraphics[width=0.98\linewidth]{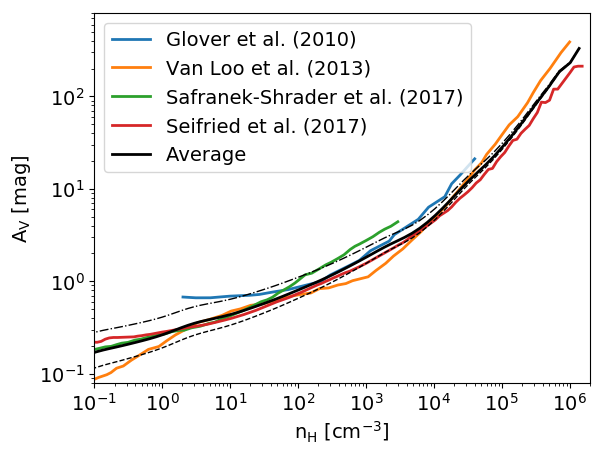}
    \caption{Relation of the local H-nucleus number density, $n_{\rm H}$, with the most probable visual extinction, $A_V$ as derived from 3D hydrodynamical models. The black solid line corresponds to the \avnh\ relation by averaging four respective relations shown. The dashed line corresponds to a lower fudge value and the dashed-dotted line to an upper fudge value as discussed in Appendix~\ref{sec:app2}.}
    \label{fig:app2}
\end{figure}

A collection of four different \avnh\ relations has been considered in this work. In particular, the relations by \citet{Glov10,VanL13,Safr17} and \citet{Seif17} have been taken into account. \citet{Glov10} (blue solid line) performed magnetohydrodynamical (MHD) models of a periodic box of side length $L=20\,{\rm pc}$ filled with a $\langle n_{\rm H}\rangle=300\,{\rm cm}^{-3}$ gas and used a six-orthogonal ray approximation to estimate the effective visual extinction at all computational elements in the box. A six-orthogonal ray approximation was also used by \citet{VanL13} (yellow solid line) to obtain the \avnh\ relation over a galactic disk with $20\,{\rm kpc}$ in diameter and a spatial resolution of $\sim8\,{\rm pc}$ or less (down to $\sim0.5\,{\rm pc}$ over a smaller, $1\,{\rm kpc}$, region to study the GMC properties). \citet{Safr17} (green solid line) performed MHD simulations of a $512\times512\times1024\,{\rm pc}^3$ box with a uniform resolution of $2\,{\rm pc}$ and they estimated the effective $A_V$ by considering the fraction between the unattenuated radiation field and the attenuated one due to dust. Finally, SILCC-Zoom simulations by \citet{Seif17} (red solid line) used a {\sc HEALPix} based \citep{Gors05} 48-ray approximation to calculate the effective $A_V$ of an $88\times78\times71\,{\rm pc}^{3}$ region at a $0.12\,{\rm pc}$ resolution. As can be seen in Fig.~\ref{fig:app2}, all aforementioned works find a very similar relation for \avnh\ for ISM regions exceeding a few tens of parsecs in size up to a whole galaxy.

The characteristic of all aforementioned \avnh\ relations is that they represent the most probable $A_V$ for a given $n_{\rm H}$ and thus they should not be considered as an one-to-one relation. As can be seen from detailed hydrodynamical models, for each $n_{\rm H}$ an \avpdf\ is associated with and needs be taken into account for a more accurate estimation of the average ISM property of that particular number density. In the case that this \avpdf\ at constant $n_{\rm H}$ is known (denoted as $\left[~\right]_{n_{\rm H}}$), the following equation (see also Eqn.~\ref{eqn:fsp2}) can be applied
\begin{eqnarray}
\langle f_{\rm sp}\rangle&=&\frac{\sum_i^q \left[f_{\rm sp,i}\right]_{n_{\rm H}}\cdot{\rm PDF}_i}{\sum_i^q{\rm PDF}_i}\\
\left[f_{\rm sp}\right]_{n_{\rm H}}&=&\frac{\sum_j^Q N_j({\rm sp})\cdot\left[{\rm PDF}_j\right]_{n_{\rm H}}}{\sum_j^Q N_j({\rm H,tot})\cdot\left[{\rm PDF}_j\right]_{n_{\rm H}}},
\end{eqnarray}
where $Q$ is the resolution (discretization) of the $\left[A_V-{\rm PDF}\right]_{n_{\rm H}}$. In this work, the above set of equations are not applied. Instead, a bounded average \avnh\ is considered as explained below.

The solid black line represents the \avnh\ relation that is assumed as the default in the present work. This is the average function when all aforementioned relations are taken into account. Since the spread in $A_V$ for constant $n_{\rm H}$ is not known, a lower and an upper value of \avnh\ is assumed as derived using the equation
\begin{eqnarray}
A_V(n_{\rm H})=g(A_V,n_{\rm H})\left\{n_{\rm H}+\lambda n_{\rm H}\left(1-\frac{\log(n_{\rm H}/n_{\rm min})}{\log(n_{\rm max}/n_{\rm min})}\right)\right\}\,\left[\frac{{\rm mag}}{{\rm cm}^{-3}}\right]
\end{eqnarray}
where $g(A_V,n{\rm H})$ is the `average' \avnh\ function (solid black line of Fig.~\ref{fig:app2}), $\lambda=1$ is a fudge factor defining the upper bound (dot-dashed line), $\lambda=-1/2$ the lower bound (dashed line) and $n_{\rm min}$, $n_{\rm max}$ are the minimum and maximum values respectively of the H-nucleus number density used in $g(A_V,n{\rm H})$. These bounds are decreasing with increasing $n_{\rm H}$ to approximate the trend of \avnh\ spread in accordance to the above hydrodynamical models.


\bsp	
\label{lastpage}

\begin{thebibliography}{99}
\bibitem[Abreu-Vicente et al.(2015)]{Abre15} Abreu-Vicente, J., Kainulainen, J., Stutz, A., Henning, T., \& Beuther, H.\ 2015, \aap, 581, A74 
\bibitem[Alves et al.(2017)]{Alve17} Alves, J., Lombardi, M., \& Lada, C.~J.\ 2017, \aap, 606, L2 
\bibitem[Balser et al.(2011)]{Bals11} Balser, D.~S., Rood, R.~T., Bania, T.~M., \& Anderson, L.~D.\ 2011, \apj, 738, 27 
\bibitem[Bell et al.(2006)]{Bell06} Bell, T.~A., Roueff, E., Viti, S., \& Williams, D.~A.\ 2006, \mnras, 371, 1865 
\bibitem[Bergin et al.(2004)]{Berg04} Bergin, E.~A., Hartmann, L.~W., Raymond, J.~C., \& Ballesteros-Paredes, J.\ 2004, \apj, 612, 921 
\bibitem[Beuther et al.(2014)]{Beut14} Beuther, H., Ragan, S.~E., Ossenkopf, V., et al.\ 2014, \aap, 571, A53 
\bibitem[Bialy \& Sternberg(2015)]{Bial15} Bialy, S., \& Sternberg, A.\ 2015, \mnras, 450, 4424 
\bibitem[Bialy et al.(2015b)]{Bial15b} Bialy, S., Sternberg, A., Lee, M.-Y., Le Petit, F., \& Roueff, E.\ 2015b, \apj, 809, 122 
\bibitem[Bialy et al.(2017)]{Bial17} Bialy, S., Burkhart, B., \& Sternberg, A.\ 2017, \apj, 843, 92 
\bibitem[Bisbas et al.(2012)]{Bisb12} Bisbas, T.~G., Bell, T.~A., Viti, S., Yates, J., \& Barlow, M.~J.\ 2012, \mnras, 427, 2100 
\bibitem[Bisbas et al.(2015a)]{Bisb15} Bisbas, T.~G., Papadopoulos, P.~P., \& Viti, S.\ 2015a, \apj, 803, 37 
\bibitem[Bisbas et al.(2015b)]{Bisb15b} Bisbas, T.~G., Haworth, T.~J., Barlow, M.~J., et al.\ 2015b, \mnras, 454, 2828 
\bibitem[Bisbas et al.(2017a)]{Bisb17} Bisbas, T.~G., van Dishoeck, E.~F., Papadopoulos, P.~P., et al.\ 2017a, \apj, 839, 90 
\bibitem[Bisbas et al.(2017b)]{Bisb17b} Bisbas, T.~G., Tanaka, K.~E.~I., Tan, J.~C., Wu, B., \& Nakamura, F.\ 2017b, \apj, 850, 23 
\bibitem[Bisbas et al.(2018)]{Bisb18} Bisbas, T.~G., Tan, J.~C., Csengeri, T., et al.\ 2018, \mnras,  
\bibitem[Black \& Dalgarno(1977)]{Blac77} Black, J.~H., \& Dalgarno, A.\ 1977, \apjs, 34, 405 
\bibitem[Bohlin et al.(1978)]{Bohl78} Bohlin, R.~C., Savage, B.~D., \& Drake, J.~F.\ 1978, \apj, 224, 132 
\bibitem[Bolatto et al.(1999)]{Bola99} Bolatto, A.~D., Jackson, J.~M., \& Ingalls, J.~G.\ 1999, \apj, 513, 275 
\bibitem[Bolatto et al.(2013)]{Bola13} Bolatto, A.~D., Wolfire, M., \& Leroy, A.~K.\ 2013, \araa, 51, 207 
\bibitem[Bournaud et al.(2015)]{Bour15} Bournaud, F., Daddi, E., Wei{\ss}, A., et al.\ 2015, \aap, 575, A56 
\bibitem[Brinch \& Hogerheijde(2010)]{Brin10} Brinch, C., \& Hogerheijde, M.~R.\ 2010, \aap, 523, A25 
\bibitem[Brunt(2015)]{Brun15} Brunt, C.~M.\ 2015, \mnras, 449, 4465 
\bibitem[Bryant \& Scoville(1996)]{Brya96} Bryant, P.~M., \& Scoville, N.~Z.\ 1996, \apj, 457, 678 
\bibitem[Butler et al.(2014)]{Butl14} Butler, M.~J., Tan, J.~C., \& Kainulainen, J.\ 2014, \apjl, 782, L30 
\bibitem[Burkhart et al.(2013)]{Burk13} Burkhart, B., Ossenkopf, V., Lazarian, A., \& Stutzki, J.\ 2013, \apj, 771, 122 
\bibitem[Burkhart et al.(2015)]{Burk15} Burkhart, B., Lee, M.-Y., Murray, C.~E., \& Stanimirovi{\'c}, S.\ 2015, \apjl, 811, L28 
\bibitem[Carniani et al.(2018)]{Carn18} Carniani, S., Maiolino, R., Smit, R., \& Amor{\'{\i}}n, R.\ 2018, \apjl, 854, L7 
\bibitem[Clark et al.(2012)]{Clar12} Clark, P.~C., Glover, S.~C.~O., Klessen, R.~S., \& Bonnell, I.~A.\ 2012, \mnras, 424, 2599 
\bibitem[Clark et al.(2013)]{Clar13} Clark, P.~C., Glover, S.~C.~O., Ragan, S.~E., Shetty, R., \& Klessen, R.~S.\ 2013, \apjl, 768, L34 
\bibitem[Dalgarno(2006)]{Dalg06} Dalgarno, A.\ 2006, Proceedings of the National Academy of Science, 103, 12269 
\bibitem[Dickman et al.(1986)]{Dick86} Dickman, R.~L., Snell, R.~L., \& Schloerb, F.~P.\ 1986, \apj, 309, 326 
\bibitem[Draine(1978)]{Drai78} Draine, B.~T.\ 1978, \apjs, 36, 595 
\bibitem[Dullemond et al.(2012)]{Dull12} Dullemond, C.~P., Juhasz, A., Pohl, A., et al.\ 2012, Astrophysics Source Code Library, ascl:1202.015 
\bibitem[Elmegreen \& Scalo(2004)]{Elme04} Elmegreen, B.~G., \& Scalo, J.\ 2004, \araa, 42, 211 
\bibitem[Federman et al.(1996)]{Fede96} Federman, S.~R., Rawlings, J.~M.~C., Taylor, S.~D., \& Williams, D.~A.\ 1996, \mnras, 279, L41 
\bibitem[Ferland et al.(1998)]{Ferl98} Ferland, G.~J., Korista, K.~T., Verner, D.~A., et al.\ 1998, \pasp, 110, 761 
\bibitem[Froebrich \& Rowles(2010)]{Froe10} Froebrich, D., \& Rowles, J.\ 2010, \mnras, 406, 1350 
\bibitem[Girichidis et al.(2014)]{Giri14} Girichidis, P., Konstandin, L., Whitworth, A.~P., \& Klessen, R.~S.\ 2014, \apj, 781, 91 
\bibitem[Girichidis et al.(2016)]{Giri16} Girichidis, P., Naab, T., Walch, S., et al.\ 2016, \apjl, 816, L19 
\bibitem[Glover \& Mac Low(2007a)]{Glov07a} Glover, S.~C.~O., \& Mac Low, M.-M.\ 2007a, \apjs, 169, 239
\bibitem[Glover \& Mac Low(2007b)]{Glov07b} Glover, S.~C.~O., \& Mac Low, M.-M.\ 2007b, \apj, 659, 1317 
\bibitem[Glover et al.(2010)]{Glov10} Glover, S.~C.~O., Federrath, C., Mac Low, M.-M., \& Klessen, R.~S.\ 2010, \mnras, 404, 2 
\bibitem[Glover \& Mac Low(2011)]{Glov11} Glover, S.~C.~O., \& Mac Low, M.-M.\ 2011, \mnras, 412, 337 
\bibitem[Glover \& Clark(2012)]{Glov12} Glover, S.~C.~O., \& Clark, P.~C.\ 2012, \mnras, 426, 377 
\bibitem[Gong et al.(2018)]{Gong18} Gong, M., Ostriker, E.~C., \& Kim, C.-G.\ 2018, \apj, 858, 16 
\bibitem[Goodman et al.(2009)]{Good09} Goodman, A.~A., Pineda, J.~E., \& Schnee, S.~L.\ 2009, \apj, 692, 91
\bibitem[G{\'o}rski et al.(2005)]{Gors05} G{\'o}rski, K.~M., Hivon, E., Banday, A.~J., et al.\ 2005, \apj, 622, 759 
\bibitem[Grassi et al.(2014)]{Grass14} Grassi, T., Bovino, S., Schleicher, D.~R.~G., et al.\ 2014, \mnras, 439, 2386 
\bibitem[Grenier et al.(2015)]{Gren15} Grenier, I.~A., Black, J.~H., \& Strong, A.~W.\ 2015, \araa, 53, 199 
\bibitem[Haworth et al.(2015)]{Hawo15} Haworth, T.~J., Harries, T.~J., Acreman, D.~M., \& Bisbas, T.~G.\ 2015, \mnras, 453, 2277 
\bibitem[Haworth et al.(2018)]{Hawo17} Haworth, T.~J., Glover, S.~C.~O., Koepferl, C.~M., Bisbas, T.~G., \& Dale, J.~E.\ 2018, \nar, 82, 1 
\bibitem[Hayden et al.(2014)]{Hayd14} Hayden, M.~R., Holtzman, J.~A., Bovy, J., et al.\ 2014, \aj, 147, 116 
\bibitem[Hennebelle \& Falgarone(2012)]{Henn12} Hennebelle, P., \& Falgarone, E.\ 2012, \aapr, 20, 55 
\bibitem[Hollenbach \& McKee(1979)]{Holl79} Hollenbach, D., \& McKee, C.~F.\ 1979, \apjs, 41, 555 
\bibitem[Hollenbach \& Tielens(1999)]{Holl99} Hollenbach, D.~J., \& Tielens, A.~G.~G.~M.\ 1999, Reviews of Modern Physics, 71, 173 
\bibitem[Hu et al.(2017)]{Hu17} Hu, C.-Y., Naab, T., Glover, S.~C.~O., Walch, S., \& Clark, P.~C.\ 2017, \mnras, 471, 2151 
\bibitem[Indriolo et al.(2015)]{Indr15} Indriolo, N., Neufeld, D.~A., Gerin, M., et al.\ 2015, \apj, 800, 40 
\bibitem[Inoue \& Inutsuka(2012)]{Inou12} Inoue, T., \& Inutsuka, S.-i.\ 2012, \apj, 759, 35 
\bibitem[Jura(1974)]{Jura74} Jura, M.\ 1974, \apj, 191, 375 
\bibitem[Kainulainen et al.(2009)]{Kain09} Kainulainen, J., Beuther, H., Henning, T., \& Plume, R.\ 2009, \aap, 508, L35 
\bibitem[Kainulainen et al.(2011)]{Kain11} Kainulainen, J., Beuther, H., Banerjee, R., Federrath, C., \& Henning, T.\ 2011, \aap, 530, A64 
\bibitem[Kim \& Ostriker(2017)]{Kim17} Kim, C.-G., \& Ostriker, E.~C.\ 2017, \apj, 846, 133 
\bibitem[K{\"o}rtgen et al.(2018)]{Kort18} K{\"o}rtgen, B., Federrath, C., \& Banerjee, R.\ 2018, \mnras,  
\bibitem[Kritsuk et al.(2011)]{Krit11} Kritsuk, A.~G., Norman, M.~L., \& Wagner, R.\ 2011, \apjl, 727, L20 
\bibitem[Krumholz(2014)]{Krum14} Krumholz, M.~R.\ 2014, \mnras, 437, 1662 
\bibitem[Krumholz \& Thompson(2007)]{Krum07} Krumholz, M.~R., \& Thompson, T.~A.\ 2007, \apj, 669, 289 
\bibitem[Langer et al.(2010)]{Lang10} Langer, W.~D., Velusamy, T., Pineda, J.~L., et al.\ 2010, \aap, 521, L17
\bibitem[Le Petit et al.(2006)]{LePe06} Le Petit, F., Nehm{\'e}, C., Le Bourlot, J., \& Roueff, E.\ 2006, \apjs, 164, 506 
\bibitem[{{Lee} {et~al.}(2015){Lee}, {Leroy}, {Schnee}, {Wong}, {Bolatto}, {Indebetouw}, \& {Rubio}}]{Lee15} {Lee}, C., {Leroy}, A.~K., {Schnee}, S., {et~al.} 2015, \mnras, 450, 2708
\bibitem[{Lee {et~al.}(2018)Lee, Leroy, Bolatto, Glover, Indebetouw, Sandstrom, \& Schruba}]{Lee18} Lee, C., Leroy, A.~K., Bolatto, A.~D., {et~al.} 2018, \mnras, 474, 4672
\bibitem[Leroy et al.(2005)]{Lero05} Leroy, A.~K., Bolatto, A.~D., Simon, J.~D., \& Blitz, L. A.\ 2005, \apj, 625, 763
\bibitem[Leroy et al.(2017)]{Lero17} Leroy, A.~K., Usero, A., Schruba, A., et al.\ 2017, \apj, 835, 217 
\bibitem[Li et al.(2018)]{Li18} Li, Q., Tan, J.~C., Christie, D., Bisbas, T.~G., \& Wu, B.\ 2018, \pasj, 70, S56 
\bibitem[Lombardi et al.(2015)]{Lomb15} Lombardi, M., Alves, J., \& Lada, C.~J.\ 2015, \aap, 576, L1 
\bibitem[Mackey et al.(2018)]{Mack18} Mackey, J., Walch, S., Seifried, D., et al.\ 2018, arXiv:1803.10367 
\bibitem[Madden et al.(2006)]{Madd06} Madden, S.~C., Galliano, F., Jones, A.~P., \& Sauvage, M.\ 2006, \aap, 446, 877 
\bibitem[Maloney \& Black(1988)]{Malo88} Maloney, P., \& Black, J.~H.\ 1988, \apj, 325, 389 
\bibitem[Maloney et al.(1996)]{Malo96} Maloney, P.~R., Hollenbach, D.~J., \& Tielens, A.~G.~G.~M.\ 1996, \apj, 466, 561 
\bibitem[McElroy et al.(2013)]{McEl13} McElroy, D., Walsh, C., Markwick, A.~J., et al.\ 2013, \aap, 550, A36 
\bibitem[Meijerink et al.(2006)]{Meij06} Meijerink, R., Spaans, M., \& Israel, F.~P.\ 2006, \apjl, 650, L103 
\bibitem[Meijerink et al.(2011)]{Meij11} Meijerink, R., Spaans, M., Loenen, A.~F., \& van der Werf, P.~P.\ 2011, \aap, 525, A119 
\bibitem[Motoyama et al.(2015)]{Moto15} Motoyama, K., Morata, O., Shang, H., Krasnopolsky, R., \& Hasegawa, T.\ 2015, \apj, 808, 46
\bibitem[Narayanan et al.(2008)]{Nara08} Narayanan, D., Cox, T.~J., Shirley, Y., et al.\ 2008, \apj, 684, 996-1008 
\bibitem[Nelson \& Langer(1997)]{Nels97} Nelson, R.~P., \& Langer, W.~D.\ 1997, \apj, 482, 796 
\bibitem[Neufeld \& Wolfire(2017)]{Neuf17} Neufeld, D.~A., \& Wolfire, M.~G.\ 2017, \apj, 845, 163 
\bibitem[Nordon \& Sternberg(2016)]{Nord16} Nordon, R., \& Sternberg, A.\ 2016, \mnras, 462, 2804 
\bibitem[Offner et al.(2013)]{Offn13} Offner, S.~S.~R., Bisbas, T.~G., Viti, S., \& Bell, T.~A.\ 2013, \apj, 770, 49 
\bibitem[Ossenkopf-Okada et al.(2016)]{Osse16} Ossenkopf-Okada, V., Csengeri, T., Schneider, N., Federrath, C., \& Klessen, R.~S.\ 2016, \aap, 590, A104 
\bibitem[Padovani et al.(2018)]{Pado18} Padovani, M., Galli, D., Ivlev, A.~V., Caselli, P., \& Ferrara, A.\ 2018, \aap, 619, A144 
\bibitem[Pak et al.(1998)]{Pak98} Pak, S., Jaffe, D.~T., van Dishoeck, E.~F., Johansson, L.~E.~B., \& Booth, R.~S.\ 1998, \apj, 498, 735 
\bibitem[Papadopoulos et al.(2004)]{Papa04} Papadopoulos, P.~P., Thi, W.-F., \& Viti, S.\ 2004, \mnras, 351, 147 
\bibitem[Papadopoulos(2010)]{Papa10} Papadopoulos, P.~P.\ 2010, \apj, 720, 226 
\bibitem[Papadopoulos et al.(2012)]{Papa12} Papadopoulos, P.~P., van der Werf, P., Xilouris, E., Isaak, K.~G., \& Gao, Y.\ 2012, \apj, 751, 10 
\bibitem[Papadopoulos et al.(2018)]{Papa18} Papadopoulos, P.~P., Bisbas, T.~G., \& Zhang, Z.\ 2018, \mnras,  
\bibitem[Pineda et al.(2017)]{Pine17} Pineda, J.~L., Langer, W.~D., Goldsmith, P.~F., et al.\ 2017, \apj, 839, 107 
\bibitem[Rachford et al.(2009)]{Rach09} Rachford, B.~L., Snow, T.~P., Destree, J.~D., et al.\ 2009, \apjs, 180, 125 
\bibitem[Requena-Torres et al.(2016)]{Requ16} Requena-Torres, M.~A., Israel, F.~P., Okada, Y., et al.\ 2016, \aap, 589, A28 
\bibitem[Richings \& Schaye(2016a)]{Rich16a} Richings, A.~J., \& Schaye, J.\ 2016a, \mnras, 458, 270 
\bibitem[Richings \& Schaye(2016b)]{Rich16b} Richings, A.~J., \& Schaye, J.\ 2016b, \mnras, 460, 2297 
\bibitem[R{\"o}llig et al.(2006)]{Roel06} R{\"o}llig, M., Ossenkopf, V., Jeyakumar, S., Stutzki, J., \& Sternberg, A.\ 2006, \aap, 451, 917
\bibitem[R{\"o}llig et al.(2007)]{Roel07} R{\"o}llig, M., Abel, N.~P., Bell, T., et al.\ 2007, \aap, 467, 187 
\bibitem[Safranek-Shrader et al.(2017)]{Safr17} Safranek-Shrader, C., Krumholz, M.~R., Kim, C.-G., et al.\ 2017, \mnras, 465, 885 
\bibitem[Schneider et al.(2015a)]{Schn15a} Schneider, N., Ossenkopf, V., Csengeri, T., et al.\ 2015a, \aap, 575, A79 
\bibitem[Schneider et al.(2015b)]{Schn15b} Schneider, N., Csengeri, T., Klessen, R.~S., et al.\ 2015b, \aap, 578, A29 
\bibitem[Schneider et al.(2016)]{Schn16} Schneider, N., Bontemps, S., Motte, F., et al.\ 2016, \aap, 587, A74 
\bibitem[Schruba et al.(2012)]{Schr12} Schruba, A., Leroy, A.~K., Walter, F., et al.\ 2012, \aj, 143, 138
\bibitem[Schruba et al.(2017)]{Schr17} Schruba, A., Leroy, A.~K., Kruijssen, J.~M.~D., et al.\ 2017, \apj, 835, 278
\bibitem[Schruba et al.(2018)]{Schr18} Schruba, A., Bialy, S., \& Sternberg, A.\ 2018, \apj, 862, 110 
\bibitem[Seifried et al.(2017)]{Seif17} Seifried, D., Walch, S., Girichidis, P., et al.\ 2017, \mnras, 472, 4797 
\bibitem[Sheffer et al.(2008)]{Shef08} Sheffer, Y., Rogers, M., Federman, S.~R., et al.\ 2008, \apj, 687, 1075-1106 
\bibitem[Smith et al.(2014)]{Smit14} Smith, R.~J., Glover, S.~C.~O., Clark, P.~C., Klessen, R.~S., \& Springel, V.\ 2014, \mnras, 441, 1628
\bibitem[Solomon et al.(1987)]{Solo87} Solomon, P.~M., Rivolo, A.~R., Barrett, J., \& Yahil, A.\ 1987, \apj, 319, 730 
\bibitem[Stark et al.(1997)]{Star97} Stark, A.~A., Bolatto, A.~D., Chamberlin, R.~A., et al.\ 1997, \apjl, 480, L59 
\bibitem[Sternberg \& Dalgarno(1995)]{Ster95} Sternberg, A., \& Dalgarno, A.\ 1995, \apjs, 99, 565 
\bibitem[Sternberg et al.(2014)]{Ster14} Sternberg, A., Le Petit, F., Roueff, E., \& Le Bourlot, J.\ 2014, \apj, 790, 10 
\bibitem[Strong et al.(2007)]{Stro07} Strong, A.~W., Moskalenko, I.~V., \& Ptuskin, V.~S.\ 2007, Annual Review of Nuclear and Particle Science, 57, 285 
\bibitem[Tafelmeyer et al.(2010)]{Tafe10} Tafelmeyer, M., Jablonka, P., Hill, V., et al.\ 2010, \aap, 524, A58 
\bibitem[Tassis et al.(2010)]{Tass10} Tassis, K., Christie, D.~A., Urban, A., et al.\ 2010, \mnras, 408, 1089 
\bibitem[Tritsis et al.(2018)]{Trit18} Tritsis, A., Yorke, H., \& Tassis, K.\ 2018, \mnras,  
\bibitem[van der Tak et al.(2007)]{vdTa07} van der Tak, F.~F.~S., Black, J.~H., Sch{\"o}ier, F.~L., Jansen, D.~J., \& van Dishoeck, E.~F.\ 2007, \aap, 468, 627 
\bibitem[van Dishoeck \& Black(1988)]{vDis88} van Dishoeck, E.~F., \& Black, J.~H.\ 1988, \apj, 334, 771 
\bibitem[van Dishoeck \& Black(1986)]{vanD86} van Dishoeck, E.~F., \& Black, J.~H.\ 1986, \apjs, 62, 109 
\bibitem[van Dishoeck(1992)]{vDis92} van Dishoeck, E.~F.\ 1992, Astrochemistry of Cosmic Phenomena, 150, 143 
\bibitem[Van Loo et al.(2013)]{VanL13} Van Loo, S., Butler, M.~J., \& Tan, J.~C.\ 2013, \apj, 764, 36 
\bibitem[Visser et al.(2009)]{Viss09} Visser, R., van Dishoeck, E.~F., \& Black, J.~H.\ 2009, \aap, 503, 323 
\bibitem[Walch et al.(2015)]{Walc15} Walch, S., Girichidis, P., Naab, T., et al.\ 2015, \mnras, 454, 238 
\bibitem[Weingartner \& Draine(2001)]{Wein01} Weingartner, J.~C., \& Draine, B.~T.\ 2001, \apj, 548, 296 
\bibitem[Wolfire et al.(1995)]{Wolf95} Wolfire, M.~G., Hollenbach, D., McKee, C.~F., Tielens, A.~G.~G.~M., \& Bakes, E.~L.~O.\ 1995, \apj, 443, 152 
\bibitem[Wolfire et al.(2003)]{Wolf03} Wolfire, M.~G., McKee, C.~F., Hollenbach, D., \& Tielens, A.~G.~G.~M.\ 2003, \apj, 587, 278 
\bibitem[Wu et al.(2015)]{Wu15} Wu, B., Van Loo, S., Tan, J.~C., \& Bruderer, S.\ 2015, \apj, 811, 56 
\bibitem[Wu et al.(2017)]{Wu17} Wu, B., Tan, J.~C., Nakamura, F., et al.\ 2017, \apj, 835, 137 
\bibitem[Xie et al.(1995)]{Xie95} Xie, T., Allen, M., \& Langer, W.~D.\ 1995, \apj, 440, 674 
\bibitem[Young \& Scoville(1991)]{Youn91} Young, J.~S., \& Scoville, N.~Z.\ 1991, \araa, 29, 581 
\end{thebibliography}
\end{document}